\documentclass[preprint2]{proto}
\usepackage{times}
\usepackage[dvipsnames]{xcolor}
\usepackage{graphicx}
\usepackage{float}
\usepackage{dblfloatfix}
\usepackage{xspace}
\usepackage{hyperref}
\usepackage[T1]{fontenc}
\usepackage{amsmath}

\usepackage{siunitx}

\definecolor{darkgreen}{HTML}{25911E}

\newcommand{\jbs}[1]{{\color{black}#1}}
\newcommand{\jb}[1]{{\color{black}#1}}
\newcommand{\tb}[1]{{\color{black}#1}}

\newcommand{\se}[1]{{\color{black}#1}}

\newcommand{\ts}{\ensuremath{t_\mathrm{stop}}\xspace}
\newcommand{\tg}{\ensuremath{t_\mathrm{grow}}\xspace}
\newcommand{\St}{\ensuremath{\mathrm{St}}\xspace}

\newcommand{\Stmax}{\ensuremath{\mathrm{St_{max}}}\xspace}
\newcommand{\rhos}{\ensuremath{\rho_\mathrm{s}}\xspace}
\newcommand{\rhog}{\ensuremath{\rho_\mathrm{g}}\xspace}
\newcommand{\Hp}{\ensuremath{H_\mathrm{p}}\xspace}
\newcommand{\Hg}{\ensuremath{H_\mathrm{g}}\xspace}
\newcommand{\Sigg}{\ensuremath{\Sigma_\mathrm{g}}\xspace}
\newcommand{\Sigd}{\ensuremath{\Sigma_\mathrm{d}}\xspace}
\newcommand{\rhod}{\ensuremath{\rho_\mathrm{d}}\xspace}
\newcommand{\vset}{\ensuremath{v_\mathrm{sett}}\xspace}
\newcommand{\vf}{\ensuremath{v_\mathrm{frag}}\xspace}
\newcommand{\amax}{\ensuremath{a_\mathrm{max}}\xspace}
\newcommand{\alphamm}{\ensuremath{\alpha_\mathrm{mm}}\xspace}
\newcommand{\Sigmap}{\ensuremath{\Sigma_{\rm p}}\xspace}
\newcommand{\cs}{\ensuremath{c_\mathrm{s}}\xspace}
\newcommand{\ecrit}{\ensuremath{\epsilon_\mathrm{crit}}\xspace}
\newcommand{\Zcrit}{\ensuremath{Z_\mathrm{crit}}\xspace}
\newcommand{\bmath}[1]{\mbox{\boldmath{$#1$}}}
\newcommand{\im}{{i\mkern1mu}}
\newcommand{\vk}{v_{\rm k}}

\title{\textbf{\LARGE COMETS AND PLANETESIMAL FORMATION}}

\author{\textbf{\large Jacob B. Simon}}
\affil{\small\em Department of Physics and Astronomy, Iowa State University, 2323 Osborn Dr., Ames, IA 50011-1026, USA}

\author{\textbf{\large J{\"u}rgen Blum}}
\affil{\small\em Institut f{\"u}r Geophysik und extraterrestrische Physik, Technische Universit{\"a}t Braunschweig, Mendelssohnstr. 3, D-38106 Braunschweig, Germany}
\author{\textbf{\large Til Birnstiel}}
\affil{\small\em University Observatory, Ludwig-Maximilians-Universit{\"a}t M{\"u}nchen, Scheinerstr. 1, 81679 Munich, Germany}
\affil{\small\em Exzellenzcluster ORIGINS, Boltzmannstr. 2, D-85748 Garching, Germany}

\author{\textbf{\large David Nesvorn\'y}}
\affil{\small\em Department of Space Studies, Southwest Research Institute, 1050 Walnut St., Suite 300, Boulder, CO 80302, USA}

\begin{abstract}
\begin{list}{ }{\rightmargin 0.in}
{\small 

In this chapter, we review the processes involved in the formation of planetesimals and comets. We will start with a description of the physics of dust grain growth and how this is mediated by gas-dust interactions in planet-forming disks.  We will then delve into the various models of planetesimal formation, describing how these planetesimals form as well as their resulting structure. In doing so, we focus on and compare two paradigms for planetesimal formation: the gravitational collapse of particle over-densities (which can be produced by a variety of mechanisms) and the growth of particles into planetesimals via collisional and gravitational coagulation. Finally, we compare the predictions from these models with data collected by the {\it Rosetta} and {\it New Horizons} missions and that obtained via observations of distant Kuiper Belt Objects.
\\~\\~\\~}%leave this in to get the correct vertical space after the abstract
\end{list}
\end{abstract}

\begin{document}

\maketitle

\section{INTRODUCTION}

The formation of planetesimals from small sub-micron grains encompasses approximately 10--12 orders of magnitude increase in scale and numerous physical processes, ranging from particle-particle sticking to angular momentum exchange between disk gas and solid particles.  While there are many unanswered questions in how exactly these physical processes work, recent years have seen substantial progress in addressing several key questions:

\begin{enumerate}
    \item How do sub-micron size grains grow to the larger mm--cm solids \jbs{(referred to as pebbles; see \ref{sec:sizedis} for a more complete definition)} that have been observationally inferred to exist in disks around young stars?
    \item How do \jbs{these pebbles} continue their growth to forming km-sized planetesimals?
    \item What are the properties of the resulting planetesimals and how do they compare with observations of Solar System planetesimal populations, such as comets and Kuiper Belt Objects (KBOs)?
\end{enumerate}

Along the theoretical front, a number of new insights have been gained.  In particular, the powerful combination of analytical techniques and numerical simulations have identified a number of instabilities \jbs{\cite[e.g.,][]{Youdin2005,Youdin2005_SGI} and mechanisms \cite[e.g.,][]{Cuzzi2008,Hopkins2016}} that can concentrate small grains into regions of sufficient density to gravitationally collapse into planetesimals.

These techniques have directly led to predictions that can be tested with current and forthcoming observational campaigns, and along this front, observational data has also been abundant. With a number missions (e.g., {\it OSIRIS-REX}, {\it Rosetta}, {\it New Horizons}) to small Solar System bodies \jbs{\cite[e.g.,][]{Stern2021}}, as well as large surveys of small body populations, such as the Outer Solar System Origins Survey (OSSOS), a plethora of new data has arrived on the physical characteristics of Solar System planetesimals \jbs{\cite[e.g.,][]{Kavelaars2021,Fraser2021}}.  Furthermore, the advent of next generation optical/IR and radio telescopes, such as the Atacama Large Millimeter/submillimeter Array (ALMA) and the Very Large Telescope (VLT) have revolutionized our understanding of both gas and small particle dynamics in planet-forming disks around nearby stars \jbs{\citep{Partnership2015,Isella2016,Huang2018}}.

Finally, laboratory experiments have been very fruitful in addressing how exactly solid particles of different compositions interact when they collide and have thus served as a strong complement to both the observational and theoretical/numerical studies \jbs{\citep{Blum2008}}. Moreover, the physical properties of planetesimals formed by different hypothesized mechanisms have also been studied in the laboratory \citep{Blum2018}, allowing a comparison to those of the most primitive bodies in the Solar System.

Indeed, the combination of theory, observations, and laboratory work has proved to be very successful in improving our understanding of planetesimal and comet formation.  In this chapter, we review such work and detail the latest understanding of how planetesimals are born.  Our approach is primarily pedagogical and as such, much of our focus will be on the physics of planetesimal formation itself, while making connections to both observations and laboratory experiments when appropriate.

\section{DUST GRAIN GROWTH}

In this section, \jbs{we describe in detail our current understanding of how small solids grow in protoplanetary disks (PPDs).  Some discussion of dust growth is discussed in Chapter 2 by Aikawa et al. However, here, we delve into the details and describe the processes that limit growth of small grains beyond mm-cm sizes, thus setting the stage for many of the mechanisms described in Section~\ref{sec:dust_to_ptsm}.} 

The evolution of solids in PPDs is influenced by many effects, but they can be broadly categorized into the two areas of 1) the evolution of their composition, size, and morphology and 2) their transport and dynamics within protoplanetary disks. It is important to realize that one cannot be treated separately from the other, because these two categories affect each other in many ways: the make-up, size and morphology of a particle affects its aerodynamical properties, and the aerodynamic properties determine how a particle is transported throughout the disk and how its environment changes along the way. The environment, in turn, determines how collisions and composition evolve: it sets the collision speeds \tb{\citep[see, e.g.,][]{Birnstiel2016}} and/or condensation/sublimation rates \tb{\citep[e.g.][]{Stammler2019}}. In the following section, we will describe how dust is transported and how it evolves through collisions and will highlight how those aspects affect each other.

We begin by introducing a crucial quantity to describe the evolution from small grains to a planetesimal: the particle stopping time. It is defined as the time scale on which the particle velocity adapts to the gas velocity through drag forces and reads

\begin{equation}
    \ts = \frac{m\,\Delta v}{F_\mathrm{drag}} = \frac{a\,\rhos}{\rhog v_\mathrm{th}},
    \label{eq:tstop}
\end{equation}
where $m$, $a$, $\rhos$, $\Delta v$, and $v_\mathrm{th}$ are the particle mass, its radius, its material density,  the absolute value of the velocity difference between gas and dust, and  the mean thermal velocity of the gas (related to the gas sound speed through a factor of order unity), respectively. The right equality in \autoref{eq:tstop} assumes the drag force to be in the Epstein regime \citep{Epstein1924,Weidenschilling1977} where the particle size is smaller than the mean-free path of the gas molecules and furthermore that the dust particles are spherical and non-fractal. Often, the stopping time is expressed by the Stokes number\footnote{\jbs{A more accurate definition of the Stokes number is discussed in \ref{sec:concentration}. However, for our current purposes, this definition will suffice.}}
\begin{equation}
    \St = \ts \cdot \Omega,
    \label{eq:stokesnum}
\end{equation}
with $\Omega$ being the Kepler frequency at the location in the disk. The Stokes number is thus a dimensionless number specifying the aerodynamic behavior of the particle. The Stokes number can therefore be seen as a ``dimensionless particle size'' or is sometimes also called the dimensionless stopping time as particles with $\mathrm{St}\ll 1$ are coupling much faster to the gas flow than the orbital time scale, while particles with $\mathrm{St} \gg 1$ require many orbits to couple.

It is this coupling by drag forces that gives rise to systematic drift motion of particles \tb{\citep{Whipple1972,Nakagawa1986}}, determines how particles are affected by turbulence \tb{\citep{Voelk1980,Ormel2007}}, and sets how strongly the radial gas flow drags particles along \tb{\citep{Takeuchi2002}}. In the following, we will discuss some of these dynamical effects in more detail.

\subsection{Vertical Settling}

\tb{A particle on an inclined orbit (and in the absence of gas) would be oscillating vertically above and below a reference plane where the vertical coordinate has its origin, which we will refer to henceforth as the disk ``mid-plane". This can be seen from the equation of motion of a particle near the mid-plane} \tb{\citep{Nakagawa1986}},

\begin{equation}
    \ddot z + z \, \Omega^2 + \frac{\dot z}{\ts} = 0,
\end{equation}
where the first two terms describe a harmonic oscillator. The remaining term is the velocity dependent drag term. The presence of the gas disk thus causes a deceleration of the particle if it moves vertically up and down through the gas, and the vertical motion of the particle can be regarded as a damped harmonic oscillator. Particles with $\mathrm{St}\ll 1$ adapt quickly to the gas velocity, and in this limit, the particles are not oscillating;  instead their velocity quickly tends to vertical gas velocity that we here assume to be zero. Due to the vertical component of gravity that accelerates the particle to the mid-plane, the particle velocity approaches a steady state between the height-dependent gravitational acceleration and the velocity dependent deceleration. Within a few stopping times, this equilibrium ($\ddot z = 0$) is reached at a terminal velocity of

\begin{equation}
    v_\mathrm{sett} = - z \, \Omega\, \St \quad (\text{for }\St\ll 1).
\end{equation}
This is analogous to dropping a feather that quickly reaches a constant velocity.

If this vertical sedimentation were to proceed unhindered, then all particles would be accumulating at the mid-plane in a razor-thin disk, similar to Saturn's rings. This is however not the case, as current scattered-light images of PPDs spectacularly depict \citep[e.g.][]{Avenhaus2018}. Some form of gas motion \tb{(likely vertically diffusive turbulence, though bulk flows are also possible)} is believed to be at play that vertically mixes the particles away from the mid-plane. In this case, a balance is reached when settling leads to dust concentration gradients that in turn lead to a diffusive flux that becomes equal, but opposing to the settling flux,
\begin{equation}
    \rhod \, \vset - D \,\rhog \frac{\partial \rhod/\rhog}{\partial z} = 0,
\end{equation}
\tb{where the diffusion flux acts to smooth out gradients in the composition as the dust is mixed by and well coupled to the turbulent gas. The solution of the above equation is a steady state dust distribution}
\begin{equation}
    \rhod(z) = \rhod(0) \frac{\rhog(z)}{\rhog(0)}\,\exp\left(\int_0^z \frac{\vset}{D} \mathrm{d}z' \right).
\end{equation}
\tb{The Schmidt number $\mathrm{Sc}$ is the ratio of gas and dust diffusivities\footnote{see the discussion in \citep{Youdin2007}}. All particles with $\St\ll 1$ are well enough coupled to the turbulent eddies of the gas, leading to a Schmidt number of effectively unity. \citet{Youdin2007} showed, that it can be approximated by $\mathrm{Sc} = 1 + \St^2$. If} the gas follows the isothermal hydrostatic solution of a Gaussian curve with scale height $H$, then around the mid-plane, the vertical density profile of particles of a given size also follows a Gaussian curve but with a scale height of
\begin{equation}
    H_{\rm p} = H \sqrt{\frac{\alpha}{\alpha + \St}}.
    \label{eq:Hp}
\end{equation}

\noindent
where $\alpha$ is a quantification of the turbulence. The history and various definitions of $\alpha$ are long and not within the scope of this chapter.  \jbs{Furthermore, $\alpha$ as used here should be distinguished from the so-called ``turbulent viscosity'' of \cite{Shakura1973}. The latter quantity depends on correlations between radial and azimuthal velocity perturbations and includes turbulent magnetic stresses.  The former is purely a dimensionless turbulent diffusion parameter, such as has been quantified by e.g., \cite{Johansen2005} and \cite{Fromang2006}.} For our purposes, it suffices to assume that the turbulence is isotropic and define the turbulent diffusion as $\alpha = \delta v^2/\cs^2$,\footnote{\jbs{More accurately, $\alpha$ should include the eddy turnover time in units of $\Omega^{-1}$. However, at the typical driving length scales for disk turbulence, the eddy turn over time is $\sim \Omega^{-1}$, making the {\it dimensionless} turnover time equal to unity.}} where $\delta v$ and $\cs$ are the turbulent velocity and sound speed, respectively.

The dust to gas ratio at the mid-plane is therefore enhanced for particles with a Stokes number $\St>\alpha$, since
\begin{equation}
    \frac{\rhod(0)}{\rhog(0)} = \frac{\Sigd}{\Sigg} \sqrt{\frac{\St}{\alpha} + 1},
\end{equation}
where $\Sigd/\Sigg$ is the ratio of the dust and gas column densities.

\subsection{\label{sec:radialdrift}Radial Drift}

A basic understanding of vertical settling now established, we can apply a similar approach for the radial and azimuthal velocity of dust and gas. The equation of motion of dust and gas are coupled by drag forces and this set of equations can be solved for a steady state, i.e. the terminal velocities, see \citet{Nakagawa1986}. Summarizing these results, the gas disk in a force balance between pressure, gravity, and centrifugal forces rotates slightly sub-Keplerian for a typical, radially decreasing pressure profile. This leads to the problem that the dust particle (that feels negligible pressure acceleration) needs to be on a Keplerian orbit to reach force balance. The resulting headwind on the particles decelerate them and remove some of their angular momentum, which in turn causes them to spiral inwards radially.  The gas, which has gained angular momentum from the particles, moves radially outward, and an equilibrium is reached, as derived by \citet{Nakagawa1986} (see also \citealp{Whipple1972} and \citealp{Weidenschilling1977}), where the dust radial velocity is

\begin{equation}
    v_\mathrm{r,dust} = - \frac{2}{(1+\epsilon)^2 \, \St^{-1} + \St}\,\eta\,v_\mathrm{K},
    \label{eq:v_drift}
\end{equation}

\noindent
$\epsilon = \rhod/\rhog$ is the dust-to-gas mass ratio,
\begin{equation}
    \eta = - \frac{1}{2} \left(\frac{H}{r}\right)^2 \frac{\partial \ln P}{\partial \ln r},
    \label{eq:eta_drift}
\end{equation}
and $P$ is the gas pressure. The resulting time scale on which particles drift inward is therefore (for $\epsilon, \St \ll 1$)
\begin{equation}
    t_\mathrm{drift} = \frac{1}{\St\,\Omega} \left(\frac{H}{r}\right)^{-2} \left|\frac{\partial \ln P}{\partial \ln r}\right|^{-1} \simeq \frac{64~\mathrm{orbits}}{\St} \, \left(\frac{H/r}{0.03}\right)^{-2},
\end{equation}
which is long compared to the orbital time scale, but short in terms of the overall disk evolution time scale. This shows that growing particles can travel substantial distances during their evolution.

Both radial drift and vertical settling are a result of the terminal velocity approximation that more generally states \citep[see][]{Youdin2005} that particles drift with respect to the gas towards higher pressure, i.e. that the dust-gas velocity difference is $-(\nabla P / \rhog)\, \ts$.

\subsection{\label{sec:st-bo-fr}Collisional Growth and Destruction}

\begin{figure*}[t!]
    \includegraphics[width=2.05\columnwidth]{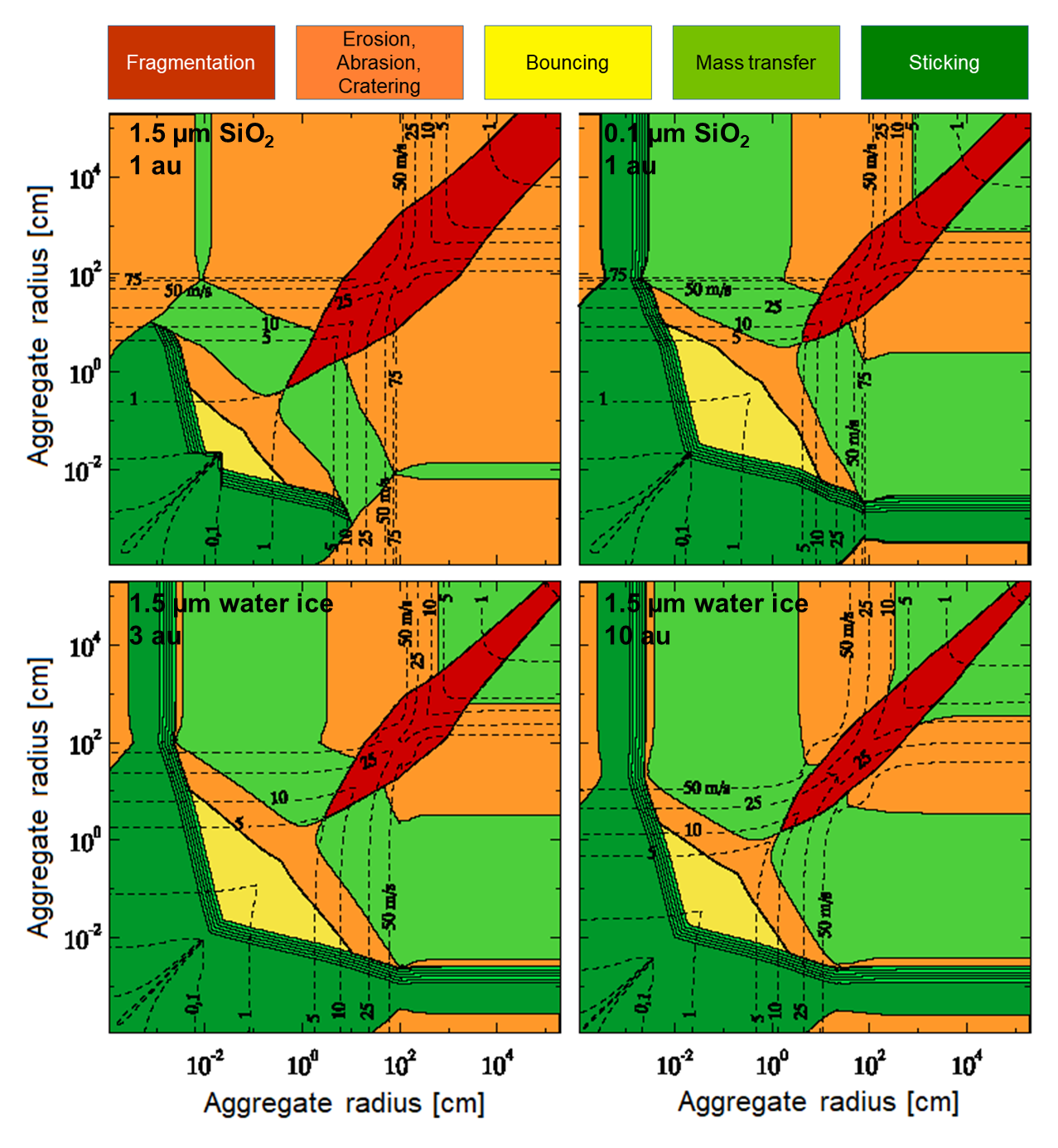}
    \caption{A dust-aggregate collision model for dust and ice aggregates after \citet{Kothe2016}. Top left: for $\mathrm{SiO_2}$ monomer grains of $1.5~\mathrm{\mu m}$ diameter at 1 au; top right: for $\mathrm{SiO_2}$ monomer grains of $0.1~\mathrm{\mu m}$ diameter at 1 au; bottom left: for water-ice monomer grains of $1.5~\mathrm{\mu m}$ diameter at 3 au; bottom right: for water-ice monomer grains of $1.5~\mathrm{\mu m}$ diameter at 10 au. The underlying PPD model is a minimum-mass solar nebula model. Image credit: Stefan Kothe.}
    \label{fig:collmod}
\end{figure*}

When two dust grains collide, their fate is determined by the collision energy as well as the material and makeup of the particles. In general, three collisional outcomes are possible: sticking, bouncing, and fragmentation, although mixtures of and transitions between these outcomes are possible too \citep[][see also Figure \ref{fig:collmod}]{Guettler2010}.

\jb{The initial grain sizes can be estimated by comparison with observations of meteorites, comets or interstellar dust. \citet{Vaccaro2015} investigated the size distribution of matrix grains in primitive carbonaceous chondrites, using high-resolution electron microscopy, and found geometric mean diameters on the order of 200~nm. \citet{Mannel2019} determined very similar mean particle sizes for the dust particles collected from comet 67P/Churyumov-Gerasimenko by the Rosetta/MIDAS instrument. Depending on the particular dust particle investigated, they found median values of the constituent grain diameters between $\sim 100$~nm and $\sim 400$~nm. In contrast to the collected samples of our own Solar System, dust grains in the interstellar medium (ISM) seem not to possess a single characteristic size value. ISM grains rather follow a power-law size-frequency distribution, with the smallest grains in the 1~nm size range and the largest ones having diameters of $\sim 500$~nm \citep{Mathis1977}. The slope of the power law is such that while most of the cross section resides in the smallest grains, most of the mass is in the largest grains. Thus, the largest effect on the mass growth will be from collisions between large grains; an appropriate range over which to investigate the collisional growth is  $0.1-1~\mathrm{\mu m}$.}

\jb{Laboratory experiments have shown that initially, dust and ice grains in PPDs generally stick upon mutual collisions \citep{Blum2008,Gundlach2015}, because their sizes are small (typically $\lesssim 1~\mathrm{\mu m}$, see above) and their collision speeds are low \citep{Weidenschilling1977,Weidenschilling1980}. Two $1~\mathrm{\mu m}$ solid silicate grains stick to each other when colliding with speeds $\lesssim 1~\mathrm{m~s^{-1}}$ \citep{Blum2008}, whereas two $1~\mathrm{\mu m}$ solid water-ice grains stick for collision velocities $\lesssim 10~\mathrm{m~s^{-1}}$ \citep{Gundlach2015}. The latter result is somewhat surprising, because the surface energy of water ice at low temperatures does not exceed that of silicates \citep{Gundlach2018,Musiolik2019}. Obviously, the surface energy is not the only material parameter with relevance to the collision and sticking behavior. \citet{Krijt2013} showed that the yield strength and the viscous relaxation time are two additional parameters that determine the collisional outcome. Recently, \citet{Arakawa2021} showed how the earlier and latest experimental data on the sticking threshold of $\mathrm{H_2O}$-ice, $\mathrm{CO_2}$-ice and $\mathrm{SiO_2}$ grains can be reconciled with the model by \citet{Krijt2013}. From the various experimental and theoretical approaches it became clear that the smaller the grains are, the higher their threshold velocity for sticking is \citep{Blum2008}. A recent study also showed that the static cohesion (a measure for the surface energy) between grains of carbon-bearing (mostly organic) substances is heavily material dependent, with the highest values for graphite and paraffin and the smallest for humic acid and brown coal; the ratio in static cohesion between these two groups is as large as a factor $1,000$ \citep{Bischoff2020}.}

Figure \ref{fig:collmod} shows examples of a state-of-the-art collision model for dust and ice grains in PPDs, based on laboratory collision experiments \citep{Kothe2016}. In the top row, the effect of monomer-grain size on the collision outcome is presented by a comparison between $\mathrm{SiO_2}$ grains with $1.5~\mathrm{\mu m}$ and $0.1~\mathrm{\mu m}$ diameter, respectively. Green colors denote the regions in which growth, either by direct sticking or by mass transfer in collisions in which one of the aggregates fragments, dominates. In the yellow (bouncing), orange (erosion, abrasion, cratering) and red regions (fragmentation), growth is not possible. It is important to note that the collision outcome not only depends on the sizes of the two colliding aggregates, but also on their collision speed, which is in the example shown determined for a minimum-mass solar nebula model (dashed contours denote collision velocities in m/s). From this comparison, it can be seen that for the more cohesive small grains, potential routes to planetesimal formation are present, namely by mass transfer in collisions between large bodies and smaller dust aggregates. However, these growth avenues are surrounded by orange and red regions so that an easy assessment of whether growth or destruction dominates is not possible. For large monomer-dust grains, the pathway to planetesimals is almost absent. Further out in the PPD, where water ice may be the dominant dust-grain material, is similar to the $0.1~\mathrm{\mu m}$ $\mathrm{SiO_2}$ grains at 1 au, even for ice particles of $1.5~\mathrm{\mu m}$ diameter (bottom row in Figure \ref{fig:collmod}). This is due to the intrinsic higher collisional stickiness of water-ice particles in comparison to the $\mathrm{SiO_2}$ grains \citep{Gundlach2015}.

\jb{It should be noted that the collision model by \citet{Kothe2016} does not explicitly include the possible change in porosity during the growth of dust agglomerates. However, an earlier model by \citet{Guettler2010} showed that there is an effect of porosity on the collisional outcome. Whenever the colliding particles are in the sticking regime, this difference is small. However, for larger aggregate masses and/or collisions speeds, compact (low-porosity) particles enter the bouncing regions, whereas fluffy (high-porosity) particles can still grow. The transition between fluffy and compact particles is usually achieved when the collision energies exceed the rolling-friction threshold, as determined by \citet{Dominik1997}. This threshold energy is required to overcome the cohesional friction against mutual rolling of the particles over their surfaces. It is apparent that a reliable modeling of dust growth has to take into account the change in porosity during the growth process. For example, \citet{Zsom2008} show that initially the dust aggregates start in extremely fluffy (or even fractal, see below) configurations, but get compacted in mutual collisions when the collision energy becomes high enough \citep{Blum2000,Weidling2009}}.

\begin{figure}[t!]
    \centering
    \includegraphics[width=\columnwidth]{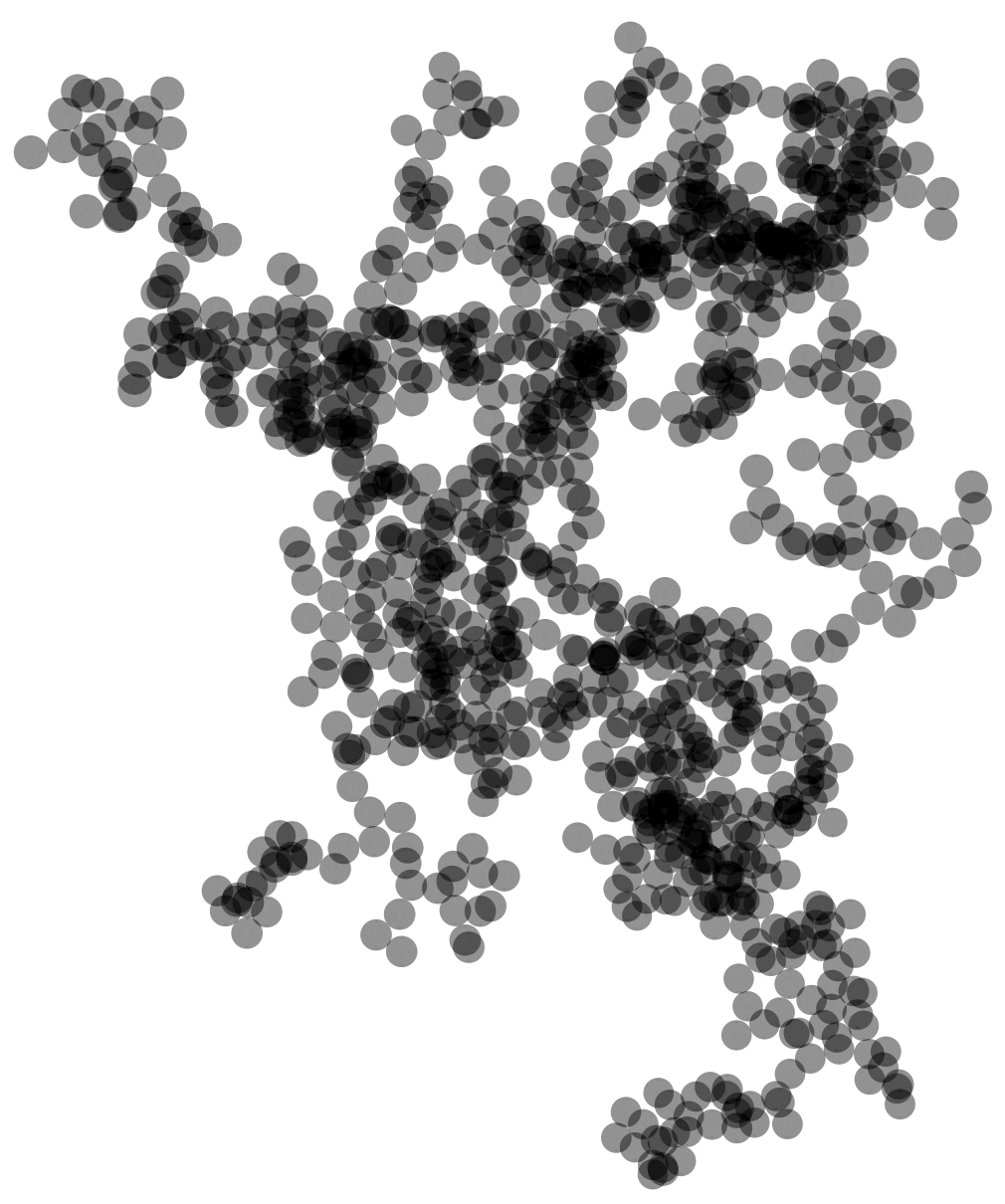}
    \caption{Example of a simulated dust aggregates with a fractal dimensions $D_\mathrm{f}=2.0$ consisting of 1,024 monomer grains with a very narrow size distribution.}
    \label{fig:fractals}
\end{figure}

Because of the initially slow collision speeds, sticking will be the dominant collisional outcome for the initial stages of grain growth in PPDs (see all four panels in Figure \ref{fig:collmod}). \citet{Dominik1997} predicted and \citet{Blum2000} experimentally confirmed that if the collision energy is much less than the energy for rolling or sliding \jb{(the latter is analogous to the rolling case and refers to the force required to slide two surfaces over each other against cohesional friction)} of the grain contacts, the collisional adhesion will be of hit-and-stick type, i.e. the grain contact will be frozen at the position of the first contact. In such a case, the growing aggregates can be described by a fractal description, $m \propto s^{D_\mathrm{f}}$, between their mass $m$ and their characteristic radius $s$, with the fractal dimension $D_\mathrm{f}$. The role of the fractal dimension can best be understood if one calculates the average mass density and the collision cross section for fractal aggregates, respectively. The former can be estimated by $\rho_\mathrm{f} \propto \frac{m}{s^3} \propto m^{1-3/D_\mathrm{f}}$, the latter by $\sigma_\mathrm{f} \propto s^2 \propto m^{2/D_\mathrm{f}}$. For $D_\mathrm{f}=3$, the aggregates behave like any solid material, i.e. $\rho_\mathrm{f} = \mathrm{const}$ and $\sigma_\mathrm{f} \propto m^{2/3}$. These aggregates are just porous, with the same porosity throughout the volume. For smaller fractal dimensions, however, e.g. $D_\mathrm{f}=2$ (see Fig. \ref{fig:fractals} for an example), the situation changes, because now $\rho_\mathrm{f} \propto m^{-1/2}$ and $\sigma_\mathrm{f} \propto m$. This means that the average density decreases with increasing aggregate mass and also from the center of mass radially outward, if one considers a specific aggregate. On the other hand, the collision cross section increases in proportion with mass and, thus, much faster than for higher fractal dimensions, which can have considerable impact on the growth rate.

The exact value of $D_\mathrm{f}$ depends on the details of the growth process and can reach from $D_\mathrm{f}\approx 1.5$ in the Brownian-motion-dominated growth regime \citep{Paszun2006} to $D_\mathrm{f}=3$ when a large particle drifts through a cloud of small grains \citep{Blum2006}. Fractal aggregates with $D_\mathrm{f}<2$ possess stopping times or Stokes numbers that are only weakly mass-dependent (see Eqs. \ref{eq:tstop} and \ref{eq:stokesnum}) and thus behave dynamically very differently than those with larger fractal dimensions. Small fractal dimensions are equivalent to preferred sticking collisions among similar-sized aggregates, i.e. the mass-frequency distribution of the growing aggregates in this case is narrow at any given time. On the other hand, for the highest $D_\mathrm{f}$ values the size distribution may be very wide or even bi-disperse \citep{Blum2006} so that the largest aggregates preferentially grow by collisions with much smaller particles. Detailed studies on this will be discussed in the following sub-section.

As the collision energy of protoplanetary dust aggregates increases with increasing mass of the particles, the criterion for hit-and-stick collisions will no longer be fulfilled and collisions will lead to the compaction of the aggregates \citep{Wada2009,Guettler2010,Zsom2010,Okuzumi2012}. Thus, dust aggregates with $D_\mathrm{f}<2$ might be converted into those with $D_\mathrm{f}>2$ and ultimately may reach $D_\mathrm{f}=3$. It should be mentioned that the compaction process depends on monomer-grain size, shape and material.

For even higher collision speeds, the colliding aggregates will either bounce or even fragment upon collision \citep{Blum2008}. While the former leads to an outside-in compaction of the aggregates \citep{Weidling2009,Guettler2010}, the latter changes the mass-frequency distribution dramatically \citep{Guettler2010,Bukhari2017}. Laboratory studies have shown that the threshold speed for fragmentation of \jb{cm}-sized aggregates \jb{of silica grains} is very close to the sticking threshold of their constituent grains \citep{Blum2008,Gundlach2015,Bukhari2017}\jb{, i.e. around $1~\mathrm{m~s^{-1}}$ for $\sim 1~\mathrm{\mu m}$ $\mathrm{SiO_2}$ grains. However, recent experiments with cm-sized aggregates consisting of micrometer-sized water-ice particles at low temperatures showed that their fragmentation speed is also at $1~\mathrm{m~s^{-1}}$ \jb{\citep{Landeck2018}} and, thus, much smaller than the sticking threshold of their constituent particles, which is around $1~\mathrm{m~s^{-1}}$ at low temperatures \citep{Gundlach2015}. In contrast to the collisional sticking process, fragmentation seems to be dominated by the surface energy. This is consistent with the findings by \citet{Gundlach2018} that the tensile strength of aggregates consisting of $\sim 1~\mathrm{\mu m}$ $\mathrm{SiO_2}$ grains and of $\sim 1~\mathrm{\mu m}$ $\mathrm{H_2O}$ grains is very similar. In conclusion of the limited empirical data available,} the survival of aggregates in collisions above a few $\mathrm{m~s^{-1}}$ is very unlikely.

In order to understand the evolution of the particle size in PPDs, given these collisional outcomes, one needs to track all possible grain-grain collisions and their resulting aggregates and fragments, a formidable task. Numerical models therefore have to simplify this evolution in one way or another. Most common approaches fall either into a mass-bin-based direct solution of the collisional evolution \citep[e.g.][]{Weidenschilling1994,Dullemond2005,Brauer2008,Birnstiel2010,Okuzumi2012} or into Monte-Carlo based approaches \citep[e.g.][]{Ormel2008,Zsom2008}, see \citet{Drazkowska2014b} for a discussion of these approaches. However, many of those detailed simulations can be understood in simpler, approximate terms following \citet{Birnstiel2012}: even in the most optimistic case, where every collision is assumed to result in perfect sticking, radial drift can usually remove particles faster than they form. A very simplified treatment of the growth time scale can be derived from monodisperse growth (e.g. \citealp{Kornet2004,Brauer2008,Birnstiel2012}) which gives
\begin{equation}
    \tg \sim \frac{1}{\epsilon\,\Omega}.
\end{equation}

\begin{figure}[t!]
    \centering
    \includegraphics[width=\hsize]{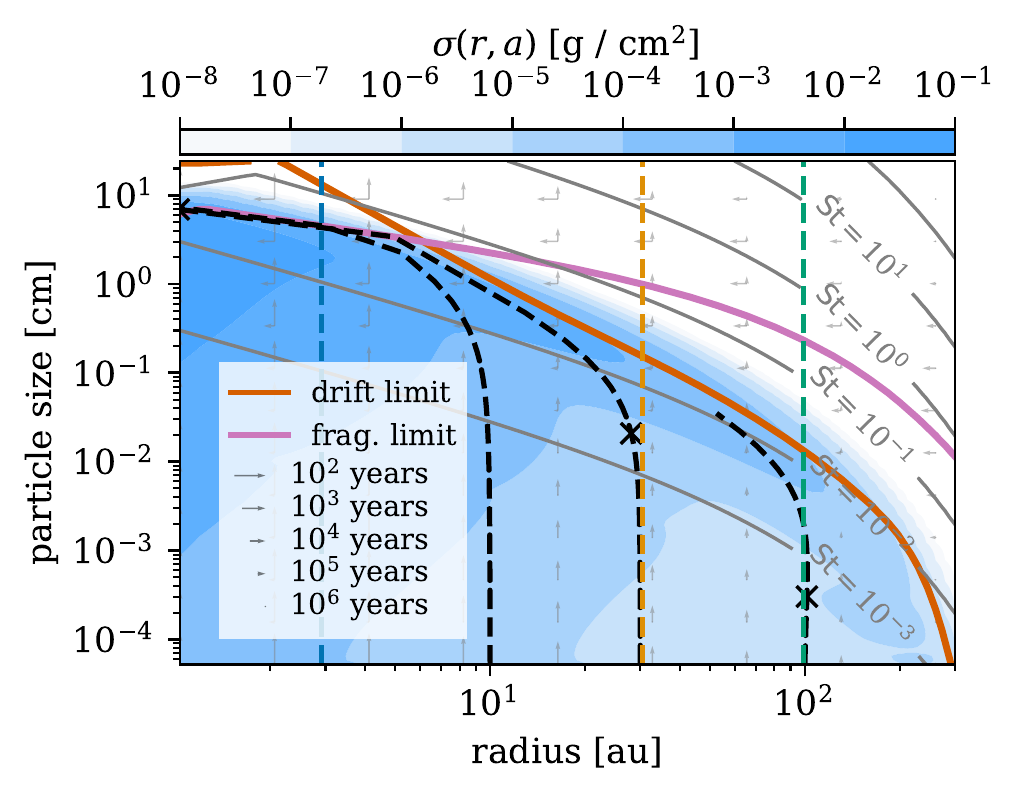}
    \caption{Simulated dust size distribution after \protect\cite{Birnstiel2010}$^1$, shown as blue contours. Gray arrows show the drift and growth rates. The growth limits are displayed as brown/orange (drift limit) and purple (fragmentation limit) lines. Gray lines denote the corresponding Stokes numbers of the particles. Monodisperse growth/drift trajectories are shown as black dashed lines, ending at \SI{700}{kyr} and denoting the position at \SI{100}{kyr} with a cross. The vertical lines correspond to the size distributions shown in \autoref{fig:dust2}. Figure adapted from \citet{Birnstiel2016}.}
    \label{fig:dust1}
\end{figure}

\noindent
This is apart a factor of a few identical with the collisional time scale. See \citealp{Birnstiel2016} for a derivation with a discussion of the approximations under which this is derived. This time scale (see the vertical arrows in \autoref{fig:dust1}) is roughly independent of particle size \citep[but see][]{Powell2019}, while the drift time scale (horizontal arrows in \autoref{fig:dust1}) does depend on particle size. Hence, small particles drift so slowly, that they mainly grow in situ, while large particles can drift inward before they substantially continue to grow. Equating these particles gives us a limiting size, called the ``drift limit'' that particles approximately can reach before drift will move them to smaller radii
\begin{equation}
    a_\mathrm{drift} \simeq 0.35\frac{\Sigd}{\rhos} \left(\frac{H}{r}\right)^{-2}\,\left|\frac{\partial \ln P}{\partial \ln r}\right|^{-1},
\end{equation}
which is shown as orange line in \autoref{fig:dust1}.

The resulting particle sizes are typically below millimeters in the outer disk but can be far beyond meters in the inner disk. This suggests that the outer disk may be limited by these dynamical effects, while in the inner disk, the above mentioned collisional effects (bouncing, erosion, fragmentation) are more likely to stop particle growth. For those disk regions, a limiting particle size can be derived if the threshold velocity for bouncing or fragmentation is compared with the size-dependent collision speeds. For turbulent velocities \citep{Ormel2007}, and e.g. a fragmentation threshold velocity \vf, the fragmentation limit becomes \citep[e.g.][]{Birnstiel2012}
\begin{equation}
    \St_\mathrm{frag} = \frac{1}{3\alpha} \, \left(\frac{\vf}{c_\mathrm{s}}\right)^2,
\end{equation}
which shows a strong dependence on the fragmentation velocity.
Similar limits can be derived for either different sources of relative velocities (e.g. drift velocity) or different threshold velocities (e.g. bouncing threshold velocity). \autoref{fig:dust1} only considers fragmentation and displays the fragmentation barrier as purple line. \autoref{fig:dust1} shows how the simulated dust size distribution (blue contours, after \citealp{Birnstiel2012}) is well contained below either the drift limit in the outer disk (orange line) or the fragmentation limit (purple line). The arrows display the approximate growth and drift rates. Particle growth is non-local in mass space. This means any particle size can interact with any other size and affect many other sizes, for example in case of a distribution of fragments. Still, the evolution of the bulk mass can often be understood by calculating trajectories in the mass/radius space (see black dashed lines in \autoref{fig:dust1}): small grains grow but do not drift significantly which results in a vertical motion in the figure. At \SI{100}{kyr} the particles have reached the position marked with a cross, which shows that by that time, particles in the outer disk have barely grown while particles in the inner disk have since long grown to large sizes and have been lost towards the inner disk. This inward radial drift can be stopped if pressure maxima are present (we explain this in detail in Section~\ref{sec:concentration} below, but also see e.g., \citealp{Pinilla2012}): at those positions, radial drift is stopped (see Eqns.~\ref{eq:v_drift} and \ref{eq:eta_drift}) and dust can accumulate.

\begin{figure}[t!]
    \centering
    \includegraphics[width=\hsize]{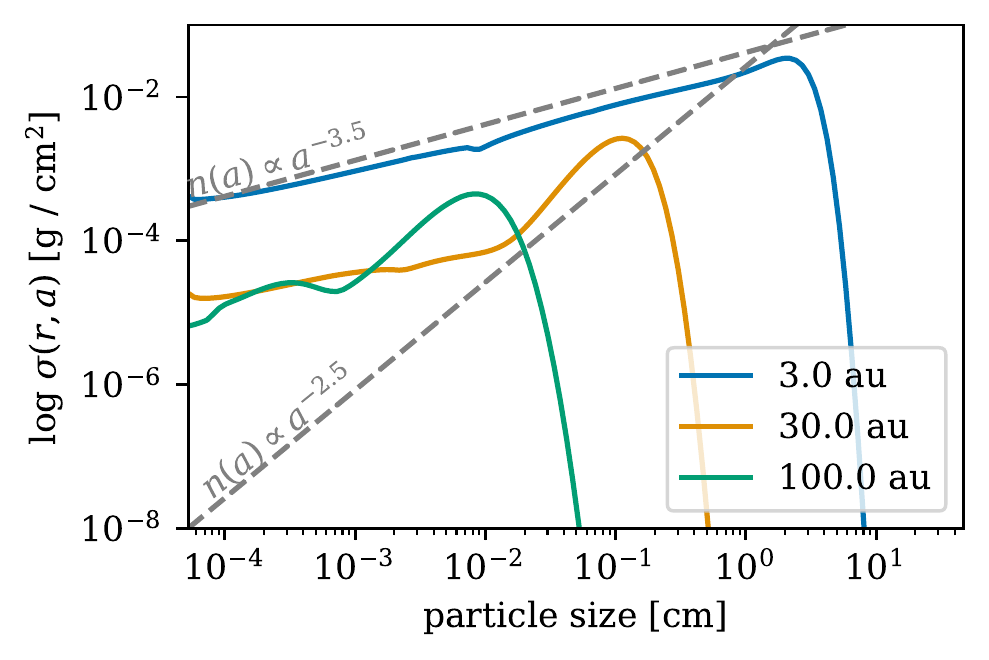}
    \caption{Dust size density distributions (surface density per logarithmic mass) measured at 3, 30, and \SI{100}{au} from the simulation in \autoref{fig:dust1}. The corresponding slope of the MRN size distribution as well as a top-heavier exponent ($\propto a^{-2.5}$) are shown as references. Figure after \citet{Birnstiel2012}.}
    \label{fig:dust2}
\end{figure}

The black-dashed trajectories in \autoref{fig:dust1} show that the particles tend to grow towards and along the drift limit, and in the inner disk along the fragmentation limit. In the latter case, continuous fragmentation and coagulation balance each other, which maintains a constant population of both small and large particles. \autoref{fig:dust2} shows the particle size distributions at different positions in the disk. It can be seen that in all cases, most of the mass resides near the largest particles. This, coupled with the fact that the velocity also scales with the particle size, means that those largest sizes carry most of the radial mass flux \citep{Birnstiel2012}. The resulting maximum particle size \amax, determines the rate at which the dust surface density is transported inward and this explains why the two-population model of \citep{Birnstiel2012} reproduces more complex simulations well.

This inward drifting, growing dust might feed dust pressure traps and/or accrete onto planets and it determines the local particle sizes that may eventually form planetesimals \jbs{\citep[][see \autoref{sec:dust_to_ptsm}]{Lambrechts2014,Lambrechts2019}}.

% \textcolor{gray}{This section will describe the physical processes involved in dust growth and fragmentation. Birnstiel will cover the theoretical modelling of this, whereas Blum will connect this to laboratory experiments. }

% \textcolor{gray}{There will be some overlap here with Chapter 8.  We have discussed this overlap with the authors of that Chapter (lead author: Poch), and we have agreed that we will describe the growth processes in detail, and they will touch on it again within the context of their chapter, but without as much detail.}

\newenvironment{orangetext}{\color{orange}}{\ignorespacesafterend}

\subsection{\label{sec:condensation}Condensational growth}

Collisional growth is not the only possible mode of grain growth in protoplanetary disks. Dust particles can also grow by deposition of vapor species on their surface. The key difference is, however, that condensation stops once the vapor is consumed, i.e. continuous condensation needs continuous resupply of condensible vapor. In contrast, collisional growth does not need to be sustained by adding mass. This means that significant growth by orders of magnitude in size can be easily achieved by collisions \citep[e.g.][]{Ormel2007b,Brauer2008,Zsom2008,Okuzumi2009,Birnstiel2010}, but via condensation would require orders of magnitude of resupplied mass. A possible way around this limitation would be continuously repeating events of sublimation and re-condensation. However, condensation/deposition is a surface effect and small particles usually dominate the dust surface area of a size distribution even if they constitute only a negligible fraction of the total dust mass. This means, that condensational growth as described can only increase the sizes of small particles efficiently \citep{Hirashita2012}. This was also seen in \citet{Stammler2017} where both coagulation and sublimation/re-condensation were simulated around the CO snow line. More recently, \citet{Ros2019} pointed out that heterogeneous ice nucleation on small silicate grains may be hindered. This, however could at most offset the problem to larger sizes: whenever there is a wide distribution of particle sizes, the smallest particles will grow more efficiently than the largest particles.

So while condensation cannot explain growth from ISM sizes to mm-cm sizes, it is still thought to be an important effect near the water snow line. As the most abundant volatile, water sublimation at the water snow line can deposit significant amounts of water vapor \citep[e.g.,]{Cuzzi2004,Ciesla2006,Garate2020}. Outward diffusion of this vapor can lead to deposition of water ice on the inward-drifting icy particles. This can locally increase both the particle size as well as the dust surface density and lead to conditions supporting the streaming instability \citep{Drazkowska2016,Schoonenberg2017}. The exact location where this takes place with respect to the snow line, depends however also on the effects of back-reaction of the dust onto the gas density distribution \citep{Hyodo2019,Garate2020}.

\subsection{\label{sec:sizedis}Grain Size Distributions}

%\textcolor{white}{\footnotemark}
%\footnotetext{The definition here follows \citet{Birnstiel2010}, where $\Sigma(r) = \int_0^{\infty} \sigma(r, a) \,\mathrm{d}\log a$.}

Starting from an idealized mono-disperse distribution of individual dust grains and assuming a high sticking probability in every occurring collision, the resulting mass-frequency distribution of the growing dust aggregates depend on (i) the velocity field among the particles and (ii) the fractal dimension of the aggregates. \jb{If we approximate the mass dependency of the velocity field by a power law $v(m) = v_0 \left(m/m_0\right)^\gamma$, with $v_0$ and $m_0$ being the velocity and mass of a monomer grain, and the collision cross section of the fractal aggregates by $\sigma = \sigma_0 \left(m/m_0\right)^{2/D_\mathrm{f}}$, with $\sigma_0$ being the collision cross section of the monomer grains, the approximate growth equation reads
\begin{equation}
    \frac{\mathrm{d}\left(m/m_0\right)(t)}{\mathrm{d}t} = \frac{1}{\tau} ~ \left(m/m_0\right)^{\gamma + (2/{D_\mathrm{f}})} ,
\end{equation}
with $\tau=1/(n_0 \sigma_0 v_0)$ and $n_0$ being the collision timescale and the initial number density of monomer grains, respectively \citep{Blum2006}.} For $\gamma + (2/D_\mathrm{f}) < 1$, we are in the regime of orderly growth, for which the average mass grows with a power law of time and the mass-frequency distribution at any given time is quasi-mono-disperse (i.e. narrow). For $\gamma + (2/D_\mathrm{f}) > 1$, the growth is characterized by the so-called runaway process in which the mass-frequency distribution becomes bi-disperse and the aggregates in the upper mass peak grow infinitely large in a finite time.

Thus, the growth behavior depends massively on the two parameters $\gamma$ and $D_\mathrm{f}$. The prior one can obtain values between $\gamma = - 1/2$ for Brownian-motion-dominated growth and $\gamma = + 1/3$ for drift-dominated velocities. The fractal dimension can obtain values between $D_\mathrm{f}=1.5-2$ for Brownian-motion-driven growth and $D_\mathrm{f}=3$ for either the pre-compaction stage (see above) or in the runaway-growth regime. Unfortunately, these two parameters are not independent and cannot easily be predicted ab initio. Thus, detailed numerical studies are required for the prediction of the growth speed, the morphology and the mass-frequency distribution of the aggregates.

As pointed out in Sect. \ref{sec:st-bo-fr}, growth stops at the bouncing barrier or at the fragmentation barrier \citep{Guettler2010,Zsom2010}. As these barriers typically occur after the sticking-with-compaction stage has been reached, the aggregates have $D=3$ and an internal porosity of $\sim 60-70 \%$ \citep{Weidling2009,Guettler2010,Zsom2010}. Their final size depends on the size and material of the monomer dust/ice grains as well as on the location in the PPD and can reach values of $\sim 1$~cm at 1 au \citep{Zsom2010} and $\sim 1$~mm at 30-50 au \citep{Lorek2018}. These aggregates have been termed "pebbles".

While growth can proceed orderly (as a well defined peak), or in a run-away fashion, the outcome of many growth (and transport) simulations tend to reach a quasi-steady-state size distribution, the shape of which is determined by the collisional outcome.

Firstly, if particles grow up to the fragmentation barrier, mass is effectively redistributed to all sizes smaller than the fragmenting particle. In that case dust mass is transported to larger sizes by growing collisions and to smaller sizes by fragmenting or eroding collisions. This quickly leads to a steady state where continuous growth and fragmentation balance each other. An example of this can be seen in \autoref{fig:dust2} at \SI{3}{au}, which shows slices of the dust distrivbution of \autoref{fig:dust1}. The resulting size distribution can depend on the way the particles fragment, particularly whether most fragmented mass is in the smallest or the largest fragments. Analytic and semi-analytic models of these scenarios were derived in \citet{Birnstiel2011}. Interestingly, for the most common regime of turbulent collision velocities and fragment distributions, the resulting size distribution is very close to $n(a) \propto a^{-3.5}$ which is the result for a fragmentation cascade, despite the different physics involved. Particles below roughly a micrometer in size, however are even more strongly depleted, such that the particle surface area is not dominated by the smallest particles present, but instead by particles of about one micron \citep{Birnstiel2018}.

Secondly, if particles do not fragment, but instead bounce off each other, most particle sizes tend to be around the bouncing barrier, leading to a very narrow size distribution with strongly reduced amounts of fine grains \citep{Zsom2010,Windmark2012b}.

Thirdly, if growth is not halted by a collisional barrier, but instead by the dynamical removal of larger particles in the drift limit, then again the particle size distribution becomes ``top-heavy'' with most mass in a narrow range around the maximum particle size, since small dust is not efficiently reproduced and instead swept up by the largest particles. The particle size distribution was measured to follow approximately $n(a) \propto a^{-2.5}$ in the drift limited case \citep[see][or \autoref{fig:dust2}]{Birnstiel2012}. The shallower slopes towards the smallest grains is caused either by radial mixing from the inner disk, where small grains are produced by fragmenting collisions (see the \SI{30}{au} case in \autoref{fig:dust2} and \citealp{Birnstiel2015}). In the \SI{100}{au} case, the small grains are remnants of the initial condition as many have not yet grown to larger sizes.

% \textcolor{gray}{This will follow immediately from the previous section: what do growth models/experiments mean for grain size distributions in the protosolar nebula? }

\subsection{Constraints from Observations}

In recent years, much progress has been made towards testing the theoretical expectations above through observations. Traditionally, this was done by measuring the spectral index at millimeter wavelengths $F_\nu \propto \nu^{\alphamm}$, as $\alphamm$ (in the optically thin Rayleigh-Jeans limit) is linked to the grain opacity $\kappa_\mathrm{abs} \propto \nu^\beta$ via $\alphamm = \beta + 2$. $\beta$ in turn depends on the maximum particle size and to a lesser extent on the particle size distribution. While particles much smaller than the wavelength typically present values of $\beta \sim 1.7$, values of $\beta < 1$ tend to be only reached if the maximum particle size is larger than the observational wavelength \citep[e.g.][and references therein]{Ricci2010,Testi2014}.

Initial disk-integrated measurements already indicated low $\beta$ values at millimeter wavelengths, indicative of millimeter sized or larger particles, which are in agreement with the predicted maximum particle sizes of theoretical models. A large caveat, however, is that the lifetime of particles of such sizes, should however be much shorter than the disk age. Theoretical models produced the right sizes, but only for a short time of a few \SI{e5}{yr} \citep{Brauer2007,Brauer2008,Birnstiel2011}. One possible solution, already envisioned by \citet{Whipple1972} was the introduction of pressure perturbations that have a vanishing pressure gradient, thus locally halting radial drift. Strong enough pressure perturbations could locally trap the large particles long enough, as required by observations \citep{Pinilla2012}.

The theoretically predicted size sorting of larger grains at smaller radii \citep{Birnstiel2012} was also confirmed by modeling the radially resolved spectral indices in \citet{Perez2012,Tazzari2016}. However, it was not until ALMA's large baseline observations became available that sub-structures were imaged \cite{Partnership2015,Andrews2016,Huang2018} that seemed to agree with the idea of pressure traps. These observations also revived the idea of \citet{Ricci2012} that a significant fraction of the flux could be optically thick -- then the observed low spectral index is not caused by large grains, but instead by optically thick emission \citep{Tripathi2017}.

ALMA has also opened the way to using polarization to measure particle sizes via self-scattering \citep{Kataoka2015}, which indicated particles not much larger than about \SI{100}{\micro m}, a clear conflict with spectral-index based measurements that indicated millimeter or centimeter sized particles. While this conflict continues to linger \citep[see][for a possible solution]{Lin2020}, the width of several of the imaged rings of DSHARP \citep{Andrews2018,Huang2018} is so narrow that dust trapping, i.e. $\St\gtrsim \alpha$ is most likely required \citep{Dullemond2018}, putting a constraint on the particle sizes. Furthermore, these rings might be providing the right conditions for planetesimal formation via the streaming instability and gravitational collapse (see \autoref{sec:SI}), as also argued for by \citet{Stammler2019} as this would naturally lead to the observed optical depths of around 0.4 in all of the well resolved sub-structures. The formation of planetesimals in pressure traps is briefly discussed further in \autoref{sec:concentration}.

\section{DUST TO PLANETESIMALS}
\label{sec:dust_to_ptsm}

Having described the formation and evolution as well as properties of small solids in PPDs, we now turn to a description of how pebbles transition to the next phase of planet formation: planetesimals.
Currently, there are two schools of thought for how planetesimals form.  The first relies on pebbles being concentrated somehow to the point where their mutual gravitational attraction overpower competing effects, such as stellar tides and turbulent diffusion. As we will see, this approach has provided the largest number of potential mechanisms, which we refer to as ``gravitational collapse models".

The second school of thought takes the opposing view of bottom-up growth.  These models, which we refer to as ``coagulation models", rely on the imperfection of the the various growth barriers described above; as described further below, there are routes toward sticking mm--cm size pebbles together to further grow in size, such as the collision-induced mass transfer from smaller aggregates to larger bodies.  Thus, coagulation models describe a route to planetesimal formation through purely collisional growth.

Most of the literature to date has focused on the former set of models, and consequently, we devote most of the remainder of this chapter to their description.  However, in Section~\ref{sec:coagmod}, we describe in detail the coagulation models.

\subsection{Gravitational Collapse Models}

As mentioned above, the gravitational collapse models all provide a route for pebbles to be concentrated sufficiently well that their mass density $\rhod$ exceeds that which is required to counteract the destructive tidal forces from the central star, the so-called Roche density,

\begin{equation}
    \rho_{\rm R} \equiv 3.5 \frac{M_*}{r^3}
    \label{eq:roche_density}
\end{equation}

\noindent
where $M_*$ is the mass of the central star, and $r$ is the radial distance of the self-gravitating object from the star (for a derivation see \citealt{Armitage2007}). 

In the remainder of this section, we describe a number of different mechanisms that all provide a route toward $\rhod \gtrsim \rho_{\rm R}$ locally, and thus for planetesimals to be born.

\subsubsection{The Goldreich-Ward Mechanism}

The first such model put forth, and in some ways, the simplest one is that of a gravitationally unstable collisionless particle disk. This mechanism, often referred to as the Goldreich-Ward mechanism \citep{Goldreich1973}\footnote{Similar work can be found in \citep{Safronov1969}.}, is no longer thought to work for reasons we will address shortly.  However, we include it here as a starting point for the discussion of other instabilities, especially as some of the ideas described here carry over to other mechanisms.

Following \cite{Binney_Tremaine}, our starting point with this discussion will be the consideration of a razor thin gaseous disk that extends infinitely, has a constant surface density $\Sigma$, has sound speed $\cs$, and is rotating uniformly at a rate $\Omega$. By introducing small perturbations of the form \jbs{${\rm exp}\left[\im (k_r r + k_\phi \phi - \omega t)\right]$ (where $\omega$ is the temporal frequency of the perturbed wave mode and $k_r$ and $k_\phi$ are the wave vector components in the radial and azimuthal directions, respectively)} to the governing equations and keeping only leading order (linear) terms (i.e., carrying out a ``linear perturbation analysis")\footnote{We encourage the reader to read \cite{Binney_Tremaine} for more details.}, one can show that the gravitational stability of this system is characterized by the so-called \citep{Toomre1964} $Q$ parameter,

\begin{equation}
    Q_{\rm g} \equiv \frac{\cs \Omega}{\pi G \Sigma}.
    \label{eq:toomre1}
\end{equation}

\noindent
\jbs{where the subscript on $Q$ denotes that it is relevant to the disk gas.}

This key parameter quantifies the ratio of stabilizing to destabilizing parameters.  More specifically, the gas disk is stabilized on small scales by thermal motions, hence the presence of $\cs$ in the numerator, whereas on larger scales, shear acts to stabilize the fluid, as shown by $\Omega$ (also in the numerator).\footnote{This dependence of stability on scales can more directly be seen by examining the dispersion relation, as shown in \cite{Binney_Tremaine}.} For more massive disks (i.e., larger $\Sigma$ in the denominator), the effect of the stabilizing parameters is diminished and the disk is more subject to gravitational instability.

More quantitatively, if $Q_{\rm g} < Q_{\rm g, crit}$, the system is gravitationally unstable, and for the setup considered here $Q_{\rm g, crit} = 0.5$ (see \citealt{Binney_Tremaine}).  However, gas is a collisional system, which is different than our collision{\it less} solid particle disk. Yet, even in this system, $Q$ remains a very useful quantity. If we define $\sigma$ as the one-dimensional velocity dispersion of particles (e.g., due to turbulent stirring by the gas) and replace $\Sigma$ with $\Sigmap$, the solid particle surface density, \jbs{our ``particle" Toomre parameter becomes}

\begin{equation}
    Q_{\rm p} \equiv \frac{\sigma \Omega}{\pi G \Sigmap}.
    \label{eq:toomre2}
\end{equation}

\noindent
\jbs{The precise value of the critical $Q$ for a differentially rotating Keplerian disk of particles (the system under consideration here), $Q_{\rm p, crit}$, will be different than the gaseous disk considered above. However, as in the gas case, $Q_{\rm p, crit} \approx  1$ and $Q_{\rm p} \lesssim 1$ implies gravitational instability of the particle layer.}

This is the crux of the Goldreich-Ward mechanism; for sufficiently large $\Sigmap$ and sufficiently low $\sigma$, $Q < 1$ and the particle layer fragments into bound objects. 

However, how viable is this mechanism given reasonable assumptions for disk parameters?
Assuming the minimum mass solar nebula model (which is arguably not the most accurate model, but serves our purposes for now; \citealt{Hayashi1981}), the gas surface density at 1~AU is $\Sigma = 1700 \mathrm{g/cm^2}$. With a standard dust-to-gas ratio of 1\%, $\Sigmap \approx 10 \mathrm{g/cm^2}$ at this location. Further assuming a solar mass star at the center of the disk, $Q = 1$ corresponds to $\sigma \approx 10~\mathrm{cm/s}$.  Since $H_{\rm p} = \sigma/\Omega$ and $\cs \sim 10^5~\mathrm{cm/s}$, we find that for gravitational instability, the ratio of particle to gas scale heights is $H_{\rm p}/H_{\rm g} \sim 10^{-4}$.

For cm-sized particles at 1 AU, $\St \sim 10^{-3}$. Using \autoref{eq:Hp}, in order to maintain such a thin particle layer, $\alpha \sim 10^{-11}$ or less, whereas even very weak turbulent stirring often leads to a minimum of $\alpha \sim 10^{-6}$--$10^{-5}$! At this point, the problem with the Goldreich-Ward mechanism should be clear; turbulence produced by any number of magnetohydrodynamic or purely hydrodynamic instabilities (see \citealt{Balbus1998} and \citealt{Lyra2019}), would {\it far} exceed this tiny number.  Even absent these processes, the velocity shear induced by the particles pushing back on the gas near the mid-plane and the slower velocity away from the mid-plane (again, due to the radial pressure gradient) leads to the Kelvin-Helmholtz instability, which {\it easily} produces turbulence that prevents such a thin particle layer from forming \citep{Cuzzi1993}.

\subsubsection{Secular Gravitational Instability}

While the Goldreich-Ward mechanism is very unlikely to produce planetesimals, there is a related phenomenon that acts similarly and {\it can} get around this thin layer problem. This process, known as the secular gravitational instability (SGI) \citep{Ward1976} comes about by considering the gravitational instability problem we just described, but with {\it gas drag included.}

The physical mechanism of this process is as follows (see also \citealt{Goodman2000}). Upon a steady state particle layer with surface density $\Sigmap$, we impose an axisymmetric particle overdensity. The width of this overdensity $\delta r$ satisfies $\Hp \ll \delta r \ll \Hg$. Due to the effects of particle self-gravity, this overdensity will contract.  However, as this contraction proceeds, the outer edge will move inward but will maintain its original angular momentum (i.e., there is no torque introduced by the self-gravity), thus particles increase their orbital velocity and resist the pull of the self-gravity. Up until now, what we have described is the stability of the system to pure gravitational instability, which can be overtaken if the mass of the overdensity is sufficiently strong.  However, another route to overpower this stability is gas drag. Since $\delta r \ll \Hg$, the gas's orbital speed is much less affected; thus, as the solid particles move inward, they feel an increased headwind, which removes their angular momentum and allows self-gravity to win and produce an even larger overdensity; from here, the whole process accelerates. Similar arguments can be applied to the inner part of the overdensity, and the result is run away growth to eventually produce gravitationally bound objects (though, there are caveats as we will see shortly)

A number of works have examined different aspects of this mechanism, ranging from its linear growth \citep{Ward2000,Youdin2005_SGI} to the behavior in the fully non-linear (i.e., no small perturbations assumed) regime (e.g., \citealt{Pierens2021}). \cite{Takahashi2014} and \cite{Takahashi2016} carried out a linear analysis of the SGI but with particle feedback on the gas included, finding that this slightly modified SGI could be responsible for producing dust rings in protoplanetary systems, such as those observed in HL Tau \citep{Partnership2015}.

Other works further explored this linear regime. \cite{Youdin2011} and \cite{Michikoshi2012} found that for small ($\leq$ mm) particles, radial drift timescales can limit the effectiveness of the SGI to regions with weak turbulent diffusion, namely $\alpha \lesssim 10^{-4}$. \cite{Shadmehri2019} found that non-axisymmetric modes of the SGI are more robust than axisymmetric modes, though are still limited by turbulent diffusion and are absent for $\alpha \ge 10^{-3}.$

More recently, a number of simulations of the non-linear SGI have been carried out. \cite{Tominaga2018} ran 2D global simulations in the $r\phi$ plane and found that indeed the SGI can form dust rings, as was hypothesized by earlier work. More recently, \cite{Pierens2021} carried out similar simulations and found that such dust rings can trigger the Rossby Wave Instability (RWI; \citealt{Lovelace1999}), which then gives rise to gravitationally bound clusters of particles via trapping in the RWI-induced vortices (see Section~\ref{sec:concentration} for more details on vortices trapping pebbles). \cite{Tominaga2020} followed up on their previous work by including radial drift and turbulent diffusion in their simulations, though in the more recent work, they assumed axisymmetry (i.e., their simulations were 1D); they found that the SGI maintained its robustness in the nonlinear regime, even with these additional effects included.

Finally, \cite{Abod2019} carried out a number of local, shearing box simulations designed to study the streaming instability (see below), but one of their simulations produced gravitationally bound clumps in the presence of gas drag but with no radial drift (which requires that the clumps form via the SGI). The formed clumps had a initial mass function very similar to the streaming instability calculations, though with a slightly steeper power law towards low masses. Furthermore, the maximum clumps that formed had masses similar that produced by the streaming instability (see below) and are thus roughly consistent with (though actually larger than) the largest Solar System planetesimals. However, as pointed out in \cite{Abod2019}, modeling the SGI in the setup that they used posed challenges, and as such, caution is warranted in interrupting these results.

Ultimately, studies of the SGI, while promising, are in their infancy and more work is needed to test the viability of this mechanism, particularly compared to others such as the streaming instability.  Indeed, since streaming-instability clumping is now known to be triggered at relatively low dust abundances (see below and \citealt{Li2021}) more work is needed to understand whether SGI could even be triggered before streaming instability clumping takes over. If the SGI can past these tests of robustness, then the next stage will be to quantify the properties of any planetesimals formed via this process.

\subsubsection{Streaming Instability}
\label{sec:SI}

The streaming instability (hereafter SI) is the most well-studied of the mechanisms to produce gravitational collapse.  As such, we will devote considerable attention to its description and the implications for planetesimal formation.

We use the term ``streaming instability" (or ``SI")  here to refer to a broad class of behaviors that have been observed in simulations that includes at the minimum two key ingredients: a background pressure gradient (and thus radial drift; see below) and momentum feedback from solids to the gas. If these criterion are included, both linear analyses \cite[e.g.,][]{Youdin2005} and non-linear simulations \cite[e.g.,][]{Johansen2007_ApJ} alike have demonstrated the unstable growth of perturbations that ultimately, as seen in the non-linear simulations, lead to clumping of solids.

However, as we describe in detail below, there has been considerable follow up work to this, which has expanded our understanding of the behavior of this coupled solid-gas system both in the linear and nonlinear state with additional affects, such as vertical gravity, turbulence, and particle self-gravity.  Some of these additional effects have led to the emergence of new routes to instability, such as the vertical shearing streaming instability of \cite{Lin2021}. We touch upon some of these extensions below, but for simplicity (and to be consistent with the literature), we collectively refer to these processes as the SI, deviating only when appropriate.

\smallskip
\begin{figure*}[t!]
    \centering
    \includegraphics[width=2.05\columnwidth]{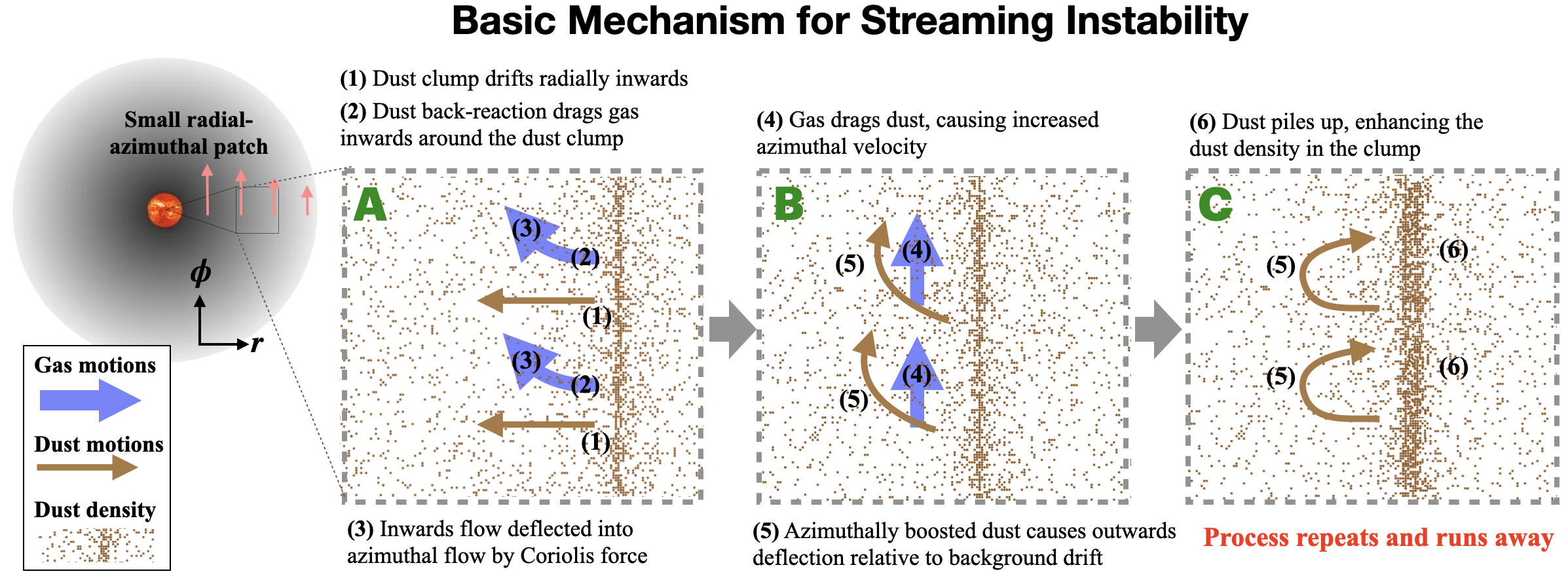}
    \caption{Basic mechanism for the streaming instability. A small dust overdensity \jbs{(Panel A)} brings gas inwards via drag. This inward gas motion is deflected azimuthally via the Coriolis force \jbs{(Panel B)}.  Dust already present downstream of the overdensity then feels an \jbs{azimuthal} boost from the deflected gas via drag, which then causes this downstream dust to move to larger radii \jbs{(as it now has increased angular momentum)}, adding to the original dust overdensity. \jbs{Note that this entire process is in a frame corresponding to the average radial drift of the dust.} \jbs{(Panel C)}. Modified from \cite{Squire2020}.}
    \label{fig:SI_mechanism}
\end{figure*}

\noindent {\it Simplest Model: Linear, Unstratified, \& No Self-gravity}\\
Our starting point is the simplest model consisting of a uniform distribution of solid particles in a uniform gas, both subject to the radial component of the central star's gravity (but not vertical!) and with angular momentum exchange between solids and gas present. By carrying out a linear perturbation analysis on the governing equations, \cite{Youdin2005} and \cite{Youdin2007} showed that this system is unstable, with growth rates and unstable wavelengths that depend on $\epsilon, \St$, and $\eta$ (as defined previously).

While a physically intuitive explanation for this instability remained elusive for many years, recent work by \cite{Squire2018, Squire2020} has provided
some clarity. This work led to the realization that in the limit of $\epsilon < 1$, the SI is actually one of a broader class of ``resonant drag instabilies''. These instabilities arise when the drift velocity of solids equal the phase velocity of a specific wave mode in the gas. Such a configuration equates to a resonance in the frame of the solids.  In the case of the SI, this resonance occurs between radial drift and epicyclic oscillations of the gas, which lead to slow growth of the dust concentration over time.

When $\epsilon > 1$, on the other hand, the feedback from the dust overwhelms the gas, which accelerates the dust concentration drastically.  Both regimes, however, operate on the same basic idea, as shown in Fig.~\ref{fig:SI_mechanism}: a dust overdensity drags gas inwards \jbs{(Panel A in the figure)}. This inward motion is deflected into the azimuthal direction via the Coriolis force \jbs{(Panel B)}.  Dust already present downstream of the overdensity then feels a boost from the deflected gas via drag, which then causes this downstream dust to move outward radially, adding to the original dust overdensity \jbs{(Panel C)}. There are a number of subtleties associated with this picture, including the difference between the $\epsilon < 1$ and $\epsilon > 1$ regimes, which are described in detail in \cite{Squire2020}. However, for the purposes of this chapter, this picture captures the essence of the SI and demonstrates how at its core, it is a concentration-inducing instability.\footnote{\jbs{While the original form of the SI, as derived by \cite{Youdin2005} is 2D in the radial-vertical plane, this diagram sacrifices information about the vertical dimension and includes the azimuthal dimension in order to provide a more intuitive understanding of the mechanism. Furthermore, the mechanism as depicted here applies to the axisymmetric case as well.}}

\medskip

\noindent {\it Nonlinear Simulations}\\
The linear phase of the SI elucidated, we now turn to a more in-depth discussion of the non-linear regime, probed with numerical simulations. While unstratified simulations have been run \citep{Johansen2007_ApJ,Bai2010_Method}, they have largely been used to test the employed code's ability to reproduce the linear growth phase.\footnote{For brevity, we will neglect a description of unstratified simulations beyond reproducing linear results.}

Most SI simulations have included the vertical component to the central star's gravity (the so-called ``stratified'' setup). Furthermore, with a few exceptions (see \citealt{Kowalik2013,Mignone2019,Schafer2020,Flock2021}), these simulations have all employed the local, shearing box approximation.  The shearing box treats a local, co-rotating patch of the disk of sufficiently small size to expand the relevant equations into Cartesian coordinates (see e.g., \citealt{Hawley1995_ApJ}). This approach is generally done when small scales ($\ll \Hg$) must be adequately resolved.

Within the context of the SI, such simulations have been done both in 2D (assuming axisymmetry) and full 3D. The primary goal of these simulations has been to study the properties of and parameters responsible for ``strong clumping'', defined here as clumping sufficiently strong so as to locally reach the Roche density.  Perhaps the most impactful of such studies have been a series of papers started by \cite{Carrera2015}; the authors of that work ran a series of 2D axisymmetric SI simulations finding an approximate ``U-shaped" region of the $\St$ vs. $Z$ (where $Z \equiv \Sigmap/\Sigma$ is the dust concentration) parameter space where strong clumping occurs.  To cover such a large parameter space, for each $\St$, $Z$ was slowly increased by removing gas from the domain.

\cite{Yang2017} expanded on this work to find that the SI can produce strong clumping at even smaller $\St$ than was determined by \cite{Carrera2015} if the numerical resolution is high enough and the simulation allowed to run for sufficiently long time. They also verified that for a subset of their parameters, the results are similar between 2D and 3D.

\begin{figure}[t!]
    \centering
    \includegraphics[width=1\columnwidth]{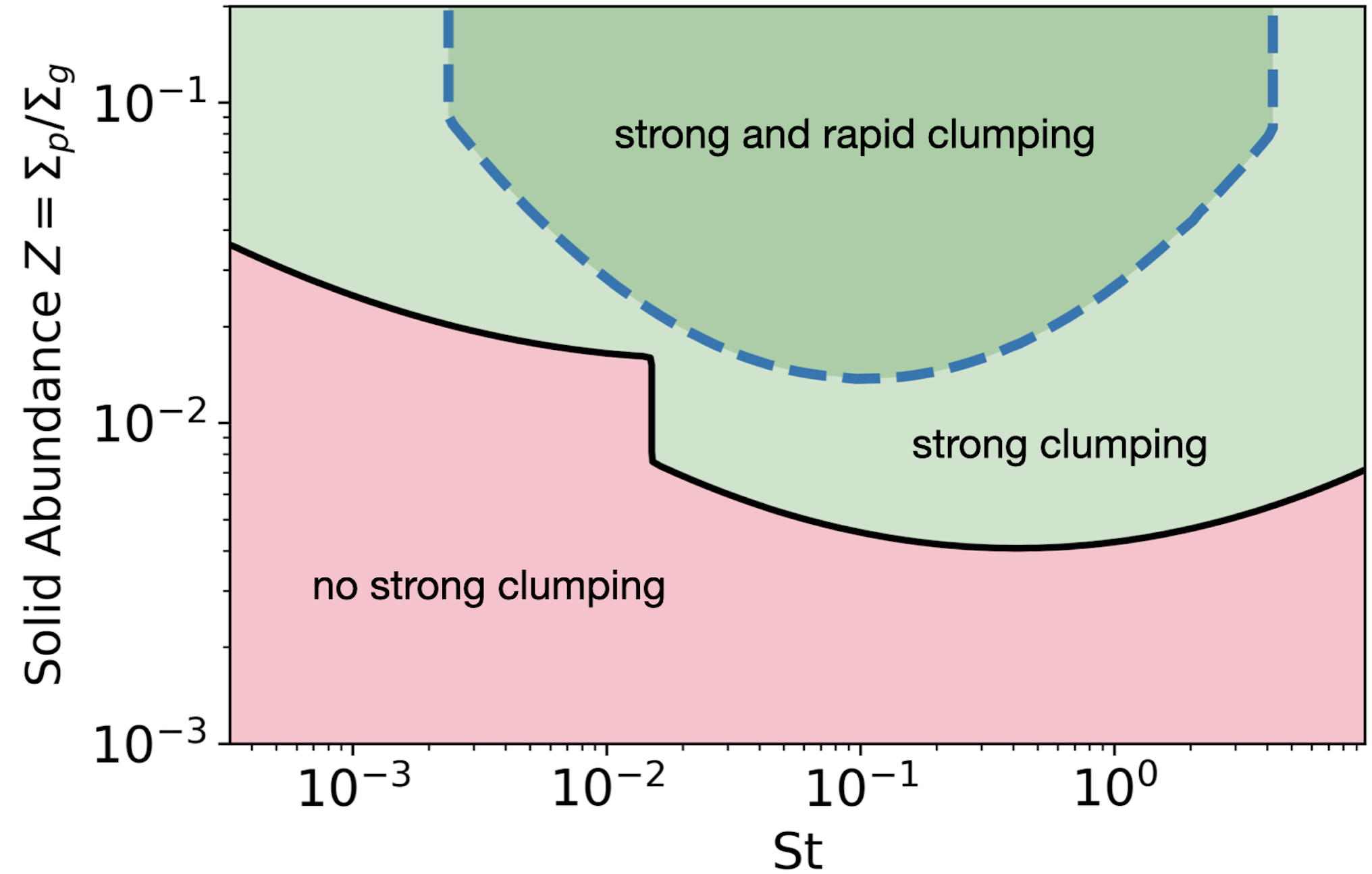}
    \caption{Regions in $\St$--$Z$ space for which strong clumping does or does not occur, based on the work by \cite{Carrera2015,Yang2017,Li2021} and assuming $\Pi = 0.05$. The pink region corresponds to no strong clumping, whereas both green regions correspond to strong clumping. The darker green region, outlined by the blue dashed line indicates the work by \cite{Carrera2015} in which strong clumping was more rapid due to the continuous removal of gas, amounting to a 50\% reduction of the remaining gas every 50 orbits of the local shearing box.}
    \label{fig:strong_clumping}
\end{figure}

More recently, \cite{Li2021} re-examined the problem again, over a larger range of $\St$ values than explored by \cite{Yang2017} but also running 2D and 3D calculations for very long integration times.  They found that the strong clumping criterion is even less restrictive than originally thought, allowing such clumping to occur for $Z = 0.004$, well below the standard value assumed to be inherited from the ISM. Figure~\ref{fig:strong_clumping} summarizes the results from these three works and indicates our current understanding of the parameters for which strong clumping is allowed.

Beyond $\St$ and $\epsilon$, the latter of which can be related to $Z$ via $Z = \epsilon \left(H_g/H_p\right)$, the third parameter that controls the linear SI growth is the pressure gradient associated with stellar illumination.\footnote{The gas pressure decreases with radius to partially balance the disk against the radial component of gravity. The remainder of the balance comes from azimuthal flow (i.e., centrifugal balance).  That pressure aids rotation in balancing gravity is the reason for the gas to be sub-Keplerian, albeit slightly.} This pressure gradient can be written in dimensionless form as

\begin{equation}
    \label{eqn:PI}
    \Pi \equiv \frac{\Delta v}{\cs}
    = - \frac{1}{2} \left( \frac{\cs}{\vk} \right) \frac{\partial \ln P}{\partial \ln r}
\end{equation}
where $\vk$ is the Keplerian velocity, $P$ is the gas pressure, and $\Delta v$ is the headwind experienced by solid particles

\begin{equation}
    \Delta v \equiv \vk - u_\phi = \eta \vk   
\end{equation}

\cite{Carrera2015} found that lower $\Pi$ decreased the threshold for strong clumping, consistent with earlier work by \cite{Bai2010_Letter}. While this third parameter may make further parameter studies of the SI expensive, recent results \citep{Sekiya2018} have demonstrated that $Z/\Pi$ is actually the more relevant parameter for determining the clumping properties of the SI.  That is, if one changes $Z$ and $\Pi$ separately but their ratio remain constant, the SI clumping appears to be very similar \citep{Sekiya2018}.

This critical result allows us to expand the parameter surveys described above to define the critical concentration parameter $Z_{\rm crit}$ for any value of $\Pi$ as (see equation 9 in \citealt{Li2021})

\begin{equation}
    \label{eqn:zcrit_pi_only}
    \Zcrit\left(\Pi\right) = \frac{\Pi}{0.05}Z_{\rm crit}\left(\Pi=0.05\right).
\end{equation}

This result has a clear physical interpretation when considering the result that larger $\Pi$ is associated with more turbulence induced by the SI \citep{Carrera2015,Abod2019}. Higher turbulence equates to a larger $H_p$, which for a given $Z$ will decrease $\epsilon$ (again because $Z = \epsilon H_g/H_p$), which as has been shown from unstratified simulations is a control parameter for the fully non-linear SI.

This result also led \cite{Li2021} to write an equation for the critical $\epsilon$ responsible for strong clumping by combining the \cite{Sekiya2018} result with their simulation suite. \cite{Li2021} also included the effect of externally driven turbulence (e.g., from the MRI) \jbs{in their equation (though external turbulence was {\it not} included in their simulations} through the traditional $\alpha$ parameter (defined above), as this also sets $H_p$ and thus $\epsilon$. While we will return to the effect of externally driven turbulence shortly, it is worth writing this critical $\epsilon$ here as it is our {\it most current} understanding of the condition for strong clumping due to the SI. From \cite{Li2021},

\begin{equation}
    {\rm log_{10}}\, \ecrit \simeq A' \left({\rm log_{10}}\,\St\right)^2 + B' \left({\rm log_{10}}\,\St\right) + C'
\end{equation}

\noindent
where

\begin{equation*}
    \begin{cases}
        A' = 0, B' = 0, C' = {\rm log_{10}(2.5)} & \text{if \,  } \St < 0.015 \\
        A' = 0.48, B' = 0.87, C' = -0.11         & \text{if  \, } \St > 0.015
    \end{cases}
\end{equation*}

This can then be included into a new equation for $\Zcrit$ that takes into account the effect of $\Pi$ (see Equation \ref{eqn:zcrit_pi_only}) and externally driven turbulence parameterized via $\alpha$,

\begin{equation}
    \label{eqn:zcrit_full}
    \Zcrit \simeq \ecrit(\St) \sqrt{\left(\frac{\Pi}{5}\right)^2 + \frac{\alpha}{(\alpha+\St)}}
\end{equation}

While the above set of equations represents our current best understanding of where in physical parameter space, strong clumping occurs and planetesimal formation is likely, numerical effects must also be quantified when employing simulations.  \cite{Yang2014} studied the effect of domain size, which is particularly important as the small scales over which the SI operate ($\ll H_{\rm g}$) restrict simulations to small domain sizes, typically $\approx 0.2H_{\rm g}$ on a side. The authors found an average separation of $\sim 0.2H_{\rm g}$ between SI-induced clumps, suggesting that at least this domain size should be used in SI simulations. The authors also found that the maximum degree of clumping (measured as the maximum particle density throughout the simulation domain) increases with increasing resolution, consistent with prior work \citep{Johansen2007_Nature,Johansen2012,Bai2010_Method}, though \cite{Yang2014} did find tentative evidence for convergence in this quantity at $160$--$320$ grid zones per $H_g$.

\cite{Li2018} followed up on this work, re-examining the effect of box size and resolution, while also exploring different vertical boundary conditions. Given the small domain size typically used in such simulations, previously used boundary conditions were either periodic or reflecting. \cite{Li2018} examined the effect of a modified open boundary (first developed for MHD simulations in \citealt{Simon2011b} and then modified by \citealt{Li2018}), finding that such a boundary led to better convergence of results with box size, a smaller $H_p$ (due to the absence of reflecting or periodic vertical gas flows), but no appreciable change in the maximum particle density.

\jbs{
More recently, \cite{Li2021} found that larger domains in the radial direction may be necessary for strong clumping when the relevant parameters lie close to the clumping boundary (the solid black line in Fig.~\ref{fig:strong_clumping}). Furthermore, \cite{Rucska2021} found evidence that scales larger than $0.2H_{\rm g}$ may be important in determining the properties of planetesimal born from the streaming instability. 
}

Clearly, a picture of what parameters lead to strong clumping is emerging.  However, there still remains more work to be done, exploring both physical parameters and numerical effects to fully flesh out the nature of strong clumping. For example, a major unanswered question is: {\it why} does strong clumping occur for the parameter choices that it does?  As pointed out by \cite{Li2021}, large linear growth rates do not necessarily lead to strong clumping. Not only does this motivate a deeper understanding of how strong clumping works, but it also warrants caution in over-interpreting the results of linear analysis. Furthermore, improving upon other conditions used in previous work, such as assuming axisymmetry for most of the $\St$, $Z$ parameter exploration, as well as neglecting particle self-gravity and its influence on strong clumping, may prove to change (yet again) the parameters for which strong clumping occurs.

\begin{figure*}[ht]
    \centering
    \includegraphics[width=2.\columnwidth]{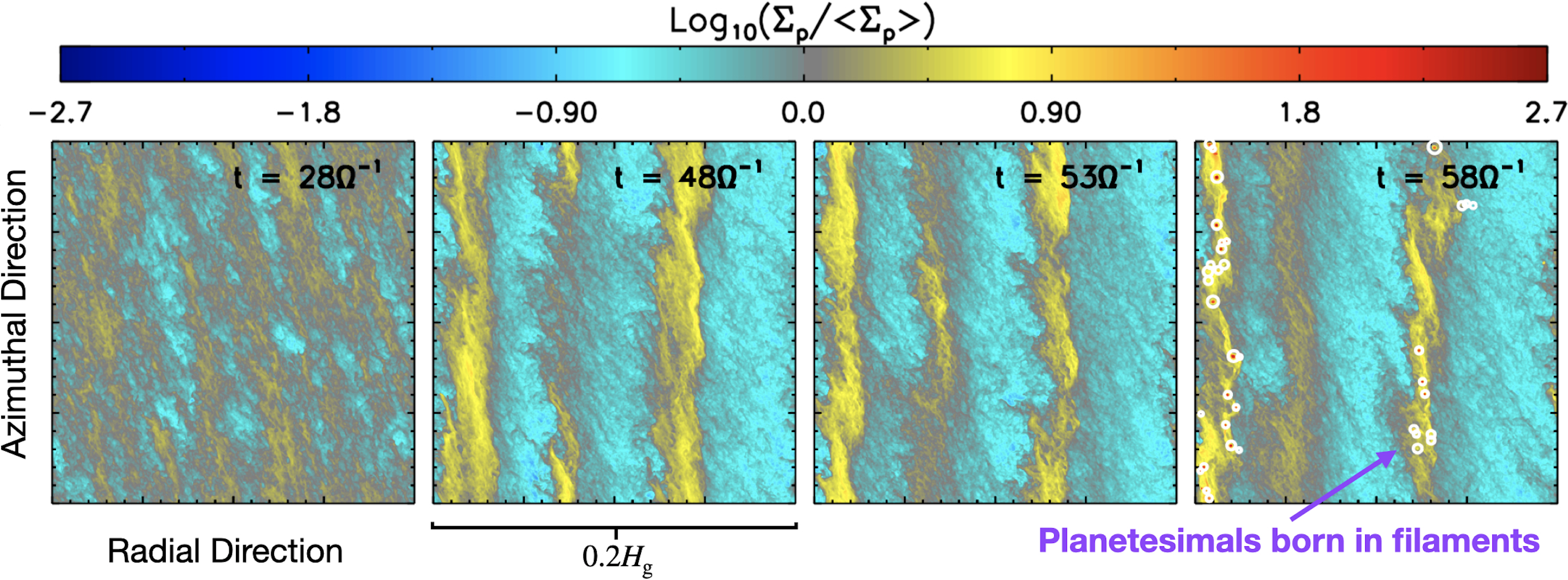}
    \caption{Log of the pebble surface density (relative to the average value) in a series of SI simulation snapshots. The simulation domain is a small patch of a disk, with the vertical axis being azimuthal and the horizontal axis radial. As the SI proceeds, pebble over-densities begin to form, ultimately creating azimuthally elongated filaments. Eventually (right-most panel) the mass density in these filaments exceeds Roche, leading to gravitationally bound clouds of pebbles. Modified from \cite{Abod2019}. \jbs{$\Omega$ as shown on the figure refers to the rotation rate at the center of the disk patch.}}
    \label{fig:SI_sigma}
\end{figure*}

\medskip

\noindent {\it Extensions: Turbulence, Multiple Particle Species}\\
There have been a number of extensions applied to the SI problem, in the linear and nonlinear regimes alike. For instance, externally driven turbulence has been shown to significantly reduce the growth rate of the SI in the linear regime (\citealt{Umurhan2020,Chen2020} and see preliminary comments on the matter in \citealt{Youdin2005}). However, as we just emphasized (and also mentioned in other works; see \citealt{Li2021}), one should be cautious in interpreting the linear growth rates as having a direct influence on the degree of strong clumping.

Indeed results from nonlinear simulations appear mixed.  Simulations by \cite{Gole2020}, for which turbulence was injected into a domain of sufficiently high resolution and small scale to resolve the SI, shows that strong clumping (and planetesimal formation; they included particle self-gravity) can be significantly reduced by turbulence.

A number of other works, on the other hand, \citep{Johansen2007_Nature,Yang2018,Schafer2020,Xu2021_Bai} included the turbulence self-consistently (e.g., included magnetic fields to activate the MRI). These studies found that the sources of turbulence explored led to solid particle concentration on scales $\sim 1$--10$\Hg$, pushing the local concentration into the streaming-unstable regime. While the obvious implication is that turbulence may actually {\it aid} in planetesimal formation, many of these studies were carried out at modest resolution. Given that the rate at which over-densities are diffused via turbulence scales with $k^2$, where $k = 2\pi/\lambda$ and $\lambda$ is the typical size scale of an overdensity, resolving smaller scales may enhance the turbulent diffusion of small-scale clumping and thus reduce or quench the production of planetesimals. Ultimately, more work remains to answer the question of whether turbulence hinders or helps SI-induced planetesimal formation; such studies are already underway.

Another extension is the inclusion of multiple particle sizes.  Of course, a non monodisperse set of particles naturally result from collisional growth/fragmentation models, as outlined above (see Sec.~\ref{sec:sizedis}).  However, for simplicity, many previous SI studies have assumed just a single size for all particles. Recently, a linear analysis carried out by \cite{Krapp2019} has shown that for $\epsilon \lesssim 1$, the fastest growth rate of the SI decreases with increasing number of discrete particle sizes without any sign of convergence (i.e., the growth rate approaches zero as the continuum limit of particle sizes is reached). This surprising result was further explored by \cite{Zhu2021} and \cite{Paardekooper2020}, which found that the multi-species SI is largely controlled by $\epsilon$ and the maximum $\St$ of the distribution, $\Stmax$.  Furthermore, for  $\epsilon \gtrsim 1$ and $\Stmax \gtrsim 1$, the SI growth rates are of order the dynamical time, regardless of the number of particle sizes.

More encouragingly, non-linear, vertically stratified SI simulations with multiple particle sizes have also demonstrated the presence of the SI \citep{Bai2010_3D,Schaffer2018} and even the convergence of particle concentration via the SI with increasing number of particle sizes \citep{Schaffer2021}. The latter work suggests that their results converge because the conditions near the disk mid-plane where the SI is active satisfy the convergent region of parameter space in the linear regime. While this result is encouraging, these considerations imply that careful tests of convergence should be carried out when running multiple-species SI simulations.

\medskip
\noindent  {\it Formation of Gravitationally Bound Clouds} \\
Once strong clumping is achieved and the Roche density surpassed, gravitational forces overpower tidal and turbulent forces (see, e.g., \citealt{Gerbig2020}) and gravitationally bound clouds form\footnote{We use the term cloud here to distinguish these diffuse, spheroidal structures from the SI-induced clumps, which are not necessarily gravitationally bound.}. This has been demonstrated numerically with a number of calculations that include particle-particle self-gravitational forces.

One such example is shown in Fig.~\ref{fig:SI_sigma}; the SI produces elongated (nearly axisymmetric) filaments that continue to increase in concentration. Eventually, the Roche density is reached locally, and gravitationally bound clouds of particles spawn off of the SI-induced filaments.  While the figure shows just one such simulation (from \citealt{Abod2019}), a number of works have demonstrated that this is a robust process \cite[e.g.,][]{Johansen2007_Nature,Johansen2015,Simon2016b,Simon2017,Abod2019,Li2019}. Furthermore, these works have shown that the largest clouds have masses on roughly the same order as the largest Solar System planetesimals and dwarf planets, such as Ceres (though as we will see shortly, this agreement is at best very approximate).

A more in-depth comparison of these clouds with observations of Solar System planetesimals is addressed in Section~\ref{sec:ploss}. However, at this point, a technical issue associated with many of the above calculations must be discussed. The algorithm to solve Poisson's equation for the particle gravitational potential, the particle-mesh (PM) method (which has been the only method employed in calculations of SI-induced planetesimal formation to date), limits the further collapse of such clouds to a factor of $10^3$--$10^4$ larger than the scales of the largest observed planetesimals ($\approx$ 100 km). The specific details of why this happens are beyond the scope of this chapter. However, this is a purely numerical limitation, and in some cases, a work-around has already been achieved. For instance, recent work with the {\sc Pencil} code has included ``sink particles'' \citep{Johansen2015}.  Another work around is an effective zoom-in on individual bound clouds with an improved treatment of gravity, as we now describe in more detail.\footnote{We should note that simulations including the entire range of scales from birth of planetesimals to final collapse are under development.}

\subsubsection{\label{sec:final}Final Stage of Collapse: Planetesimals are Born}

Recently, \cite{Nesvorny2021} carried out a series of N-body calculations with a modified version of {\sc Pkdgrav} that solves particle gravity with a tree algorithm (see \citealt{Barnes1986}), includes collisions between particles, and properly accounts for energy dissipation via a new superparticle method (see \citealt{Nesvorny2020}). Each of these calculations focused on individual particle clouds extracted from SI calculations.  Furthermore, the particle collisions are expected to be inelastic: they extract the kinetic energy from the collapsing cloud, which allows it to collapse further towards denser and denser objects.

\begin{figure*}[ht]
    \centering
    \includegraphics[width=2.\columnwidth]{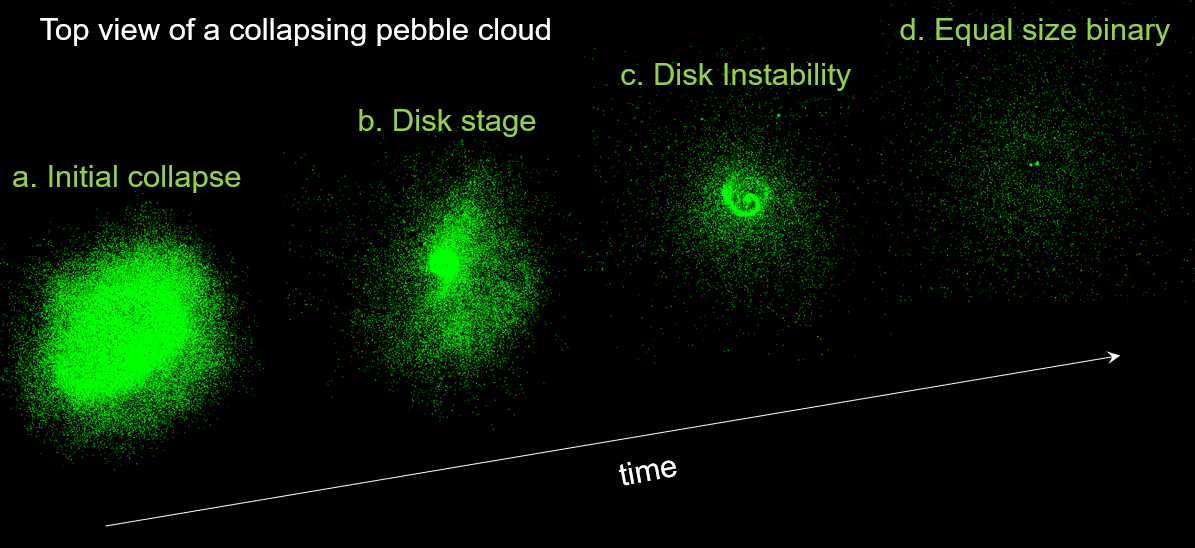}
    \caption{Gravitational collapse of a particle cloud. The four snapshots show different stages of collapse at simulation time
        $t=0$, 6, 13 and 50 yr (from left to right). The simulation was performed with the superparticle module of the {\tt PKDGRAV} $N$-body code (Nesvorn\'y et al. 2020). The view is projected down the clump angular momentum axis.
        Each frame is 450,000 km across. The equal-size binary in the right-hand panel has primary diameter $D_1 \approx 110$ km, secondary diameter $D_2 \approx 78$ km, semimajor axis $a_{\rm b}=10400$ km, and eccentricity $e_{\rm b}=0.19$. \jbs{The binary represents 98\% of the initial mass and 58\% of the initial angular momentum of its parent cloud. }
    }
    \label{fig:cloud_collapse}
\end{figure*}

\cite{Nesvorny2021} found that the particle clouds rapidly collapse into short-lived disk structures from which
planetesimals form (see Fig.~\ref{fig:cloud_collapse}). The planetesimal properties depend on the cloud's scaled angular momentum,
$l=L/(M R_{\rm H}^2 \Omega)$, where $L$ and $M$ are the angular momentum and mass, $R_{\rm H}$ is
the Hill radius \jbs{(i.e., the radius at which the gravity of the planetary body or planetesimal dominates over that from the Sun)}, and $\Omega$ is the orbital frequency. Low-$l$ particle clouds produce tight
 (or contact) binaries and single planetesimals. Compact high-$l$ clouds give birth to binary
planetesimals with attributes that closely resemble the equally sized binaries found in the
Kuiper Belt (see Section~\ref{sec:ploss}). Furthermore, 50\%-100\% of the original mass of a particle cloud is converted to
the final largest planetesimal or planetesimal binary. The remaining $<50$\% of the original mass
remains in particles and small planetesimals \citep{Robinson2020}.

We will return shortly to binary production in collapsing particle clouds, particularly within the context of comparison with observations.  However, at this point, it is worth describing studies of interior planetesimal structures, as this will also be directly relevant to the examination of actual Solar System bodies.

\medskip

\noindent {\it Planetesimal Structures - Predictions from Theory}\\
\cite{WahlbergJansson2014}, and the subsequent works of these authors (see below) considered gravitational collapse of pebble clouds, though their setup was generic (i.e., not taken from an SI calculation) and had $L \sim 0$ (i.e., no global rotation). These authors developed statistical schemes to characterize the effects of collisions on the pebble size
distribution, with different collisional regimes being considered, including sticking, bouncing, mass
transfer, and fragmentation of pebbles \citep{Guettler2010}. They found that the initial mass
of a pebble cloud determines the interior structure of the resulting planetesimal.

The pebble collision speeds in low-mass clouds are below the threshold for fragmentation ($\sim 1-10 ~ \mathrm{m~s^{-1}}$, see Sect. \ref{sec:st-bo-fr}), forming
pebble-pile planetesimals consisting of the primordial pebbles from the PPD.
Planetesimals above 100 km in radius should consist of mixtures of dust
(pebble fragments) and pebbles which have undergone substantial collisions with dust and
other pebbles. If comet-sized planetesimals form by gravitational collapse of small pebble
clouds, this model would predict that the recently visited Solar System bodies 67P/Churyumov-Gerasimenko and Arrokoth (by {\it Rosetta} and {\it New Horizons}, respectively) are pebble-pile  planetesimals consisting of primordial pebbles from the solar nebula.  If, instead, comets are collisional fragments of much larger planetesimals, the memory of original pebbles may
have been lost.

These results were extended \citep{WahlbergJansson2017b} by allowing the individual
pebble sub-clouds to contract at different rates and including the effect of gas drag (on the
contraction speed and energy dissipation). Their results yield comets that are porous pebble-piles
with particle sizes varying with depth. The interior should consist of primordial pebbles with
a narrow size distribution, yielding higher porosity. The surface layers should be a mixture
of primordial pebbles and pebble fragments. The detailed predictions depend on the applied
fragmentation model \citep{WahlbergJansson2017a}.

% If figures is needed, consider the right panel of figure 5 in Wahlberg Jansson \& Johansen
% (2017) to illustrate the above paragraph. There is also figure 3 from our Nature Astronomy paper.  

The physical properties of primordial pebble piles (ignoring pebble fragmentation during collapse) were calculated by \citet{Skorov2012} and \citet{Blum2018} and compared to known properties of comets (see Sect. \ref{sec:ploss}). \citet{Skorov2012} calculated the tensile strength of the surface layers of pebble piles, which is many orders of magnitude lower than that of non-hierarchical, porous bodies consisting of (sub-)micrometer-sized dust/ice grains. The reason for the low strength is that the number of grain-grain contacts, which determines the cohesion of granular matter under low-gravity conditions, is greatly reduced. This in turn results from the contacts occurring only in the pebble-pebble contact area, which is much smaller than the pebble cross section. Other major differences to the non-hierarchical makeup of planetesimals are the vastly different heat conductivity and gas permeability, which are important properties to understand the thermal evolution of planetesimals and the activity of comets (see below).

Due to the hierarchical setup of these kinds of planetesimals (dust/ice grains -- pebbles -- planetesimal), a number of characteristics are present:\\
\textit{Low mass density / high porosity.} Due to the effect of bouncing after the pebbles had been formed in the PPD, their internal volume filling factor was estimated to be in the range $\Phi_\mathrm{int}\approx 0.33-0.58$ \citep{Guettler2009,ORourke2020}. The pebbles themselves form a pebble pile in the gravitational collapse that has a filling factor of  $\Phi_\mathrm{pp}\approx 0.55-0.64$ \citep{ORourke2020}. Thus, small planetesimals ($\lesssim 100$~km) as a whole should possess a volume filling factor of $\Phi_\mathrm{pl}=\Phi_\mathrm{int} \cdot \Phi_\mathrm{pp} \approx 0.18-0.37$ and a mass density of $\rho_\mathrm{pl} = \Phi_\mathrm{pl} \cdot \rho_\mathrm{solid}$, with $\rho_\mathrm{solid}$ being the average mass density of the solid monomer grains. For large planetesimals ($\gtrsim 100$~km), their self-gravity is so large that the pebbles collapse and fill the macroscopic void spaces. Thus, these planetesimals have reduced porosities and increased densities and also possess other very different physical properties \citep[for details see][]{Blum2018}.\\
\textit{Low tensile, compressive and fragmentation strengths.} Due to the macroscopic size of the pebbles, the inner cohesion of planetesimals formed in a gentle gravitational collapse of a pebble cloud is extremely weak \citep{Skorov2012}. For mm- to cm-sized pebbles, tensile strength values for the planetesimals of $\lesssim 1$~Pa are expected. Similar low values are expected for the compressive strength \citep{ORourke2020}. In contrast, a homogeneous (i.e. non-hierarchical) packing of microscopic solid grains leads to tensile strengths $\gtrsim 1$~kPa \citep{Blum2018}. 

\jb{For the dynamical evolution of planetesimals, it is essential to consider their collisional behavior. A characteristic value often used in collisional evolution studies is the fragmentation strength, which is the collision energy per unit mass required to yield a largest fragment with half the mass of the original body. \citet{Katsuragi2017} and \citet{sansebastian2020} showed that the fragmentation strength scales with the static tensile strength. Thus, the energy required for the fragmentation of pebble-pile planetesimals can be many orders of magnitude smaller than in the non-hierarchical case \citep{Krivov2018b,sansebastian2020}.}
\textit{ High gas permeability.} The macroscopic void spaces between the pebbles in a planetesimal that was formed by a gentle gravitational collapse of a pebble cloud  play an important role in vapor transport \citep[see][for the influence of porosity on heating and gas transport]{Lichtenberg2016}. Besides current Solar System bodies exhibiting gas and/or dust activity, gas transport might be important in the first few million years after their formation if the abundance of short-lived radiogenic nuclei is high enough to vaporize volatile (e.g. water ice) or super-volatile (e.g. CO ice) materials \jb{\citep{Lichtenberg2021,Golabek2021,Malamud2022}}. Due to a negative temperature gradient in the radial direction \jbs{(in this context, radial means away from the center of the planetesimal)}, volatiles could be driven radially outward and could lead to their enhancement close to the surface of the planetesimal.\\
\textit{Low solid-state and high radiative thermal conductivity.} Besides the gas permeability and the mechanical strengths, the heat conductivity is heavily dependent on the size of the building blocks of the planetesimal. For a homogeneous body consisting of microscopic dust or ice grains, the heat transport occurs through the network of grains. Thus, the heat conductivity decreases for increasing porosity and increasing grain size. In contrast, a pebble-pile makeup of the planetesimals may lead to an increased heat transport, due to the influence of radiative heat transfer, in particular at higher temperatures and for larger pebble sizes \citep{bischoff2021}. The strong temperature dependency of the heat conductivity in the pebble case can lead to interesting diurnal effects, with high energy transport from the surface into the sub-surface regions during periods of high-intensity insolation and low energy transport at night times \citep{bischoff2021}.

As with the other subsections in Section \ref{sec:dust_to_ptsm}, we will return to planetesimal/comet structures shortly as we make a comparison with observations.  However, for now, we conclude this section with a brief discussion of another route to producing planetesimals via gravitational collapse.

\subsubsection{Concentration in Gas Features}
\label{sec:concentration}

The final gravitational collapse mechanisms we will discuss are the broad group of ``concentration mechanisms". These mechanisms provide a route (alternative to the SI) toward concentrating pebbles to the point that their local density surpasses the Roche density. In what follows, we only briefly describe these mechanisms, as the literature is considerably less extensive compared with that of the SI.

The simplest concentration mechanism is an axisymmetric gas pressure enhancement. These can be induced by any number of things, including magnetohydrodynamical effects \cite[e.g.,][]{Johansen2009c,Simon2014,Bai2014,Bethune2017}, hydrodynamic structures from instabilities (e.g, the vertical shear instability; \citealt{Schafer2020}), \jbs{and planets themselves, which produce pressure maxima at the outer edge of planet-induced gaps \cite[e.g.,][]{Picogna2015,Fedele2017,Hendler2018,Baruteau2019,Perez2019,Veronesi2020}}. Indeed, ALMA has observed the effect of dust concentration in such rings in a number of planet-forming disks \citep{Huang2018}, though their origin has not yet been well constrained by such observations.

The process by which such concentration occurs is as follows. Pebbles on the outer part of a pressure enhancement drift towards increasing pressure due to the headwind they feel from the partially pressure supported sub-Keplerian gas (see Section~\ref{sec:radialdrift}). However, on the other side of the pressure enhancement (i.e., at smaller radii), the opposite is true. In order for the pressure enhancement to be long lived, the Coriolis force and gas pressure gradient must balance (this is known as geostrophic balance). Thus, inward of the enhancement, gas is moving super Keplerian, which provides a tailwind to pebbles, increasing their angular momentum; the pebbles move towards pressure maxima and are trapped in the case where ${\rm d}P/{\rm d}r \ge 0$.\footnote{In the case in which a local maximum is not produced, the inward radial drift of pebbles continues, but their drift slows down as they enter the pressure enhancement causing a ``traffic jam''.}

% I want a more intuitive explanation for the following paragraph for the final version
A related process is the concentration of pebbles in vortices.  Gas vortices are another form of geostrophic balance, but without axisymmetry.  In the non-inertial co-rotating shearing box frame, cyclonic vorticies (that is with ${\bmath \nabla} \times {\bmath v} > 0$), are sheared out by the anti-cyclonic shear flow. Only anti-cyclonic vortices survive, and in this  configuration a high pressure region exists within the vortex that balances the Coriolis force; geostrophy is maintained. 

\jbs{As the vortex orbits the central body (at Keplerian velocities), it sweeps up pebbles that are inwardly drifting from larger radii.  These pebbles are captured in the outer region of the vortex and are swept along with the vortex streamlines.}
However, since they experience drag but no pressure gradient, they move within the vortex at a higher velocity relative to the gas, which causes them to slowly spiral towards the center of the vortex. \jbs{An apt analogy is to think of the vortex as a miniature disk with the role of radial gravity being replaced by the Coriolis force.  Thus the pebbles orbit the center of the vortex as they would a disk, at a higher velocity than the gas.  The resulting headwind on the pebbles causes them to spiral inward to the center of the vortex. }

\jbs{
Indeed, there is already evidence that dust can become entrapped by these vortices. For instance, early dust continuum observations with ALMA \cite[e.g.,][]{Marel2013,Perez2014} depicted the presence of large-scale asymmetric concentrations of dust grains, which has been interpreted as trapping of these grains in vortices. 
}

\jbs{
A third mechanism, known as {\it turbulent clustering} is the concentration of pebbles due to turbulent gas in planet-forming disks \citep{Klahr1997,Cuzzi2008,Pan2011,Hopkins2016,Hartlep2017,Hartlep2020}, namely through the interaction between pebbles and turbulent eddies. Within this mechanism, pebbles that are coupled to the turbulent gas are centrifugally ``flung" outward away from the centers of turbulent vorticies/eddies and are concentrated between the eddies to the point of gravitational collapse.  However, for this to work effectively, several conditions must be satisfied: 

\begin{enumerate}
    \item The pebble clump is of sufficiently high density to resist the destruction by fluid motion within the turbulent eddy as the clump and eddy interact
    \item The pebble clump is of sufficiently high density that it can resist disruption from ``ram pressure" (i.e., feeling a headwind as it moves through the gas)
    \item The pebble clump is of sufficiently {\it low} density that the feedback from pebbles does not destroy the turbulent eddies.
\end{enumerate}

More quantitatively, if one considers pebbles with a specific stopping time $t_{\rm stop}$, the Stokes number at length-scale $\ell$ is defined as\footnote{\jbs{The ratio of stopping time to eddy turnover time is the {\it true} definition of the Stokes number and the quantity St defined in Equation~\ref{eq:stokesnum} is the Stokes number at scale $H$ in which $t_{{\rm eddy},H} = \Omega^{-1}$. When planetary scientists and astronomers use the term Stokes number, they most often to refer to the stopping time in units of $\Omega^{-1}$, which is equivalent to the Stokes number at the driving scale of the turbulence.} } 

\begin{equation}
    {\rm St}_\ell \equiv \frac{t_{\rm stop}}{t_{{\rm eddy},\ell}}
\end{equation}

\noindent
where $t_{{\rm eddy},\ell}$ is the eddy turnover time at scale $\ell$. For scales smaller than $H$, $t_{{\rm eddy},\ell} < \Omega^{-1}$.

In the case of a turbulent cascade of energy toward smaller scales (e.g., Kolmogorov turbulence; \citealt{Kolmogorov1941}), $t_{\rm eddy}$ decreases with decreasing spatial scale $\ell$. Thus, for a given pebble size, ${\rm St}_\ell$ increases with decreasing $\ell$. 

The necessary conditions for collapse, as enumerated above, are best reached at scales where ${\rm St}_\ell \sim 0.3$ \citep{Hartlep2017}. At larger scales, the Stokes number is sufficiently small that the pebbles are strongly coupled to turbulent eddies and thus cannot be concentrated to meet the above collapse criteria.  On smaller scales, the pebbles are only very weakly coupled to the gas and thus feel the turbulent motions as kicks that do not appreciably change the pebble density. The ``sweet spot" for planetesimal formation happens at intermediate values, namely ${\rm St}_\ell \sim 0.3$ (see \citealt{Hartlep2017} and \citealt{Hartlep2020}).

We will return to this mechanism later in this chapter within the context of predictions for planetesimal sizes.  For now, however, it is worth mentioning that while this mechanism is promising, more work is needed to study its role in planetesimal formation and in particular in comparing predictions made by this model with Solar System constraints.} 

Finally, it is essential to point out that these processes are not mutually exclusive with the SI.  Indeed, it has recently been shown that pebble concentration \jbs{in axisymmetric pressure bumps \citep{Taki2016,Onishi2017,Carrera2021,Carrera2021_Weak,Xu2021_Bai,Xu2022} can provide a sufficient increase in localized dust-to-gas ratio such that the SI becomes active near the peak of the pressure bump.  However, while this process may be robust for some parameters (e.g., the strength of the force generating the bumps; \citealt{Carrera2021_Weak}), the simulations by \cite{Carrera2022} suggest that for sufficiently small pebbles, planetesimals are only formed when these pebbles are completely trapped in the pressure bump. In this case, which requires a bump of sufficiently high amplitude that it becomes unstable to the RWI (based on the conditions in \citealt{Ono2016}), planetesimals are only formed by concentrating the pebbles beyond the Roche density; planetesimals were formed from direct gravitational collapse with no help from the SI.  

Similarly, vortices can also provide sufficient concentration to kickstart the SI. In particular, \cite{Raettig2015} and \cite{Raettig2021} carried out local simulations of pebble concentration in vortices and found that the SI can be activated for dust-to-gas ratios much less than 1\%. 
}

\jbs{Ultimately, however, these additional mechanisms coupled with the SI deserve further study. Encouragingly, such efforts are already underway.}

\subsection{\label{sec:coagmod}Coagulation Models}

Although the pebble-collapse model has had a great impact in the scientific community and can explain many observational facts (see below), there are alternative planetesimal-formation models that still merit study. In this section, we discuss the ``coagulation models", of which there are two different types. The first involves the growth (via sticking) of particles beyond the various barriers discussed above, whereas the second involves the gravitational growth of small planetesimals into larger ones.  We discuss both here for completeness, though given the limited literature in this area compared to collapse models, this section is necessarily short.

\subsubsection{Growth Past the Barriers}

The obstacles of the bouncing and fragmentation barriers discussed in Section \ref{sec:st-bo-fr} are not absolute. The physical principles to overcome these barriers have been discussed before \citep[see, e.g.,][]{Blum2018} and will only be briefly reviewed here.

The first possibility is to avoid reaching these barriers altogether. \citet{Okuzumi2012} and \citet{Kataoka2013} pointed out that extremely small (i.e., nanometer-sized) monomer grains possess a very high sticking-bouncing threshold \citep[see Fig. 13 in][]{Blum2008} and a \jb{rather high} tensile (cohesive) strength \citep[see Fig. 4 in ][]{Gundlach2018}. This means that the growing aggregates
may remain in the fractal growth regime (see Sect. \ref{sec:sizedis}) long enough
to grow to macroscopic sizes. \citet{Okuzumi2012} and \citet{Kataoka2013} show that aggregates can grow to cm-sizes with a low fractal dimension $D_\mathrm{f} \lesssim 2$ so that their gas-grain response time does not change by much and radial drift is slow. Although the fractal aggregates are very rigid, their mechanical stiffness is finite. Thus, collisions among cm-sized fractal aggregates lead to a compaction without changing the sticking efficiency. \citet{Kataoka2013} also took into account compaction through gas drag and self-gravity. Beyond the collisional-compaction size, the aggregates continue to grow with only slightly increasing densities (from $\sim 10^{-5}-10^{-4}~\mathrm{g~cm^{-3}}$ for $\sim$cm-sized aggregates to $\sim 10^{-3}~\mathrm{g~cm^{-3}}$ for $\sim$100-m-sized bodies). Self-gravity is responsible for a rather steep increase in density for bodies above $\sim$100 m in size. Ultimately, the resulting planetesimals have sizes of $\sim$10 km and densities in the range $\sim 0.1-1~\mathrm{g~cm^{-3}}$. With this growth pattern, the radial-drift barrier (see Sect. \ref{sec:radialdrift}) can be avoided. \jb{Recently, \citet{Kobayashi2021} showed that based on such a scenario even the cores of the giant planets can form in a relatively short time.} Although the basic assumptions of high sticking threshold velocity and high tensile strength is certainly undoubted for very small monomer grains, the collision behavior of the extremely fluffy bodies above $\sim$1 cm in size \se{cannot easily be tested in the lab and thus remains uncertain}. In particular, ignoring fragmentation as a possible collision outcome, deserves a re-visit in future lab experiments.
%Thus, it is unclear how realistic the simulations by \citet{Okuzumi2012} and \citet{Kataoka2013} are.

%In terms of binarity and spin state, the model by \citet{Kataoka2013} does not give answers, because it treats sticking events or gravitational encounters only statistically. 

\jbs{A second possibility is that of mass transfer during collisions.} Although fragmentation in collisions among dust aggregates in principle leads to the mass loss of the colliding bodies, there are situations for which this is not the case for one of the collision partners. If the mass ratio of the two aggregates is large and the impact energy is moderate, only the smaller aggregate will fragment and may transfer some of its mass to the intact target aggregate \citep{Meisner2013}. Based upon the mass-transfer process, \citet{Windmark2012a,Windmark2012b,Windmark2012c,Garaud2013} showed that planetesimal formation is feasible, in spite of the dominance of fragmentation in collisions among similar-sized bodies. \citet{Windmark2012a,Windmark2012b,Windmark2012c} based their dust-evolution model on the collision model by \citet{Guettler2010} and took into account a variety of collisional outcomes. In order to overcome the bouncing barrier and enter the mass-transfer regime, \citet{Windmark2012a} artificially inserted 1-cm-sized seed particles, large enough to start the mass-transfer process. In contrast,  \citet{Windmark2012b,Windmark2012c} used velocity-probability distribution functions to initiate the mass-transfer process. While \citet{Windmark2012a} needed $10^6$ years to grow planetesimals with 100 m size, \citet{Windmark2012b,Windmark2012c} achieved this result in $\sim 5 \cdot 10^4$ years. \citet{Garaud2013} used a similar approach as \citet{Windmark2012b,Windmark2012c} and achieved aggregate sizes on the order of several hundred meters within $10^4$ years at 1 au, whereas further out, the aggregates remained much smaller.
%As in (i), the models utilizing the mass-transfer process to grow planetesimals cannot predict the occurrence of binaries or the spin-state of their final bodies.
Of critical importance for the credibility of the models is under which conditions the mass-transfer growth process dominates over mass-loss effects, such as fragmentation, cratering, or erosion. As reviewed by \citet{Blum2018} and based upon the dust-aggregate collision model by \citet{Guettler2010} with later supplements from lab experiments, fragmentation is important as long as the size ratio between the colliding bodies is $\lesssim 10$ under solar-nebula conditions; cratering becomes the dominating process when the projectile sizes are $\gtrsim 1$~cm, because in that case the impact energies are too large for mass transfer to occur; finally, erosion happens when the projectiles are $\lesssim 100~\mathrm{\mu m}$ in size \citep{Schraepler2018}. This leaves only a very small size window for the projectiles to reach the mass-transfer realm. Whether fragmentation, cratering, and erosion can completely prevent the formation of planetesimals by the mass-transfer process needs still to be clarified.

\jb{In conclusion, more empirical data on the collision behavior of fluffy aggregates are required to determine whether protoplanetary growth of solid bodies by coagulation inevitably stop at pebble sizes or continue towards planetesimal scales.}

\subsubsection{Formation of 100 km Planetesimals}

The second type of coagulation model has been examined in a number of works \cite[e.g.,][]{Stern1997,Weidenschilling1997,Kenyon2004,Kenyon2012,Schlichting2011,Schlichting2013} and is built on the assumption that smaller grains have already grown to some distribution of initial sizes, often $\sim$~km.  These smaller planetesimals then collisionally grow to larger 10-100~km size bodies.

The basic physics of how this works is relatively straightforward. At the scales associated with the initial planetesimal populations (e.g., km scales), $\St \ggg 1$, and drag forces are negligible.  Instead, small bodies grow larger when they collide with one another, ``sticking" together primarily due to gravitational forces.  Gravitational focusing also plays a role, though is significantly more relevant for larger bodies as the square of the impact parameter is enhanced by a number that is linearly proportional to the mass (assuming equal mass bodies; see \citealt{Armitage2007}).

Again, this mechanism for planetesimal formation\footnote{Though, one could argue that because these models often {\it start} with planetesimals, albeit small ones, that it is more a mechanism to change the size distribution of planetesimals than it is a process to form them in the first place.} has not been as extensively studied as the gravitational collapse models. However, later in this chapter, we will come back to it and the predictions that it makes compared with the SI model.

\section{\label{sec:ploss}PLANETESIMALS IN THE OUTER SOLAR SYSTEM}

Having discussed the various mechanisms thought to produce planetesimals, we now turn to a specific discussion of comets and Kuiper Belt Objects (KBOs).  We focus in particular on comparing observed and/or measured characteristics of KBOs and comets to those predicted by models.

\subsection{Mass and Size Distributions}
\label{sec:size_dist}

The size distribution of planetesimals in the Solar System has been one of the primary diagnostics employed to test planetesimal formation theory.  In this section, we first describe the size distribution inferred from planetesimal formation calculations, followed by a discussion of the impact of collisional evolution, ending with a comparison with observational data (a brief discussion on this can also be found in the chapter by Fraser et al.).

\subsubsection{Primordial Mass Distribution}
\label{sec:primordial_mass_dist}

As discussed previously, a number of works have explored the characteristics of SI-generated planetesimals, generally agreeing that the largest planetesimals have masses roughly consistent with large Solar System planetesimals.  In this section, we dive deeper into this topic:  the initial mass and (equivalently) size distributions of planetesimals.  As described above, many SI simulations prevent collapse below certain scales due to numerical limitations. Thus, in what follows, the size we reference for gravitationally bound pebble clouds is that which the total mass bound within a cloud would have if it were collapsed to material densities, generally assumed to be $\sim 1 {\rm~g~cm^{-3}}$.

\cite{Johansen2015} carried out high resolution simulations of SI-induced planetesimal formation in order to allow a large range of planetesimal sizes to form. Modeling the mass distribution of planetesimals as a simple power law,

\begin{equation}
    \label{dist_sp}
    N(>M_p) \propto M_p^{-p+1},
\end{equation}

\noindent
they found that $p \approx 1.6$.  More recently, a number of works followed up on this and found that $p \approx 1.6$ independent of relative strength of gravity (effectively the value of $Q$), numerical resolution \citep{Simon2016b}, $\St$, $Z$ \citep{Simon2017}, and $\Pi$ \citep{Abod2019}; that is, with this simple power law shape to the mass and size distributions, the so-called ``slope" appeared to be nearly universal.

While \cite{Abod2019} did fit a simple power law to their planetesimal masses, they were able to increase the number of detected bound clouds with a tree-based algorithm based on halo finders in cosmological simulations (the {\sc Plan} code; \citealt{Li2019_PLAN}). They found that an exponentially tapered distribution,

\begin{equation}
    \label{dist_exp1}
    N(>M_p) \propto M_p^{-p+1} \exp \left[-M_p/M_c\right],
\end{equation}

\noindent
where $M_c$ is a characteristic ``cut-off'' mass, fit the planetesimal masses significantly better than the simple power law. In their fit, $p \approx 1.3$, approximately independent of $\Pi$ (other parameters were not explored in this work).

This work was roughly consistent with \cite{Johansen2015} and \cite{Schafer2017}, though both studies fit a slightly modified version of the exponentially tapered distribution, introducing a third parameter, $\beta$:

\begin{equation}
    \label{dist_exp2}
    N(>M_p) \propto M_p^{-p+1} \exp \left[\left(-M_p/M_c\right)^{\beta}\right].
\end{equation}

\noindent
Given their resolution limits, \cite{Schafer2017} were unable to constrain $p$ but found that $\beta = 0.3$--0.4, independent of their domain size. \cite{Johansen2015} carried out simulations in a smaller domain and at higher resolution (in terms of grid cells per $\Hg$). They found best fit values $p = 1.6$ and $\beta = 4/3$; the slope of the power law component was consistent with \cite{Simon2016b} and \cite{Simon2017} but the extent of exponential tapering was significantly larger than \cite{Schaefer2017}.\footnote{We reserve discussion of the remaining parameter, $M_c$, to Section~\ref{sec:size_dist_compare} below.}

Most recently, \cite{Li2019} used {\sc Plan} and the highest resolution SI simulations to date to further quantify the mass distribution.  Analyzing two simulations with different $\St$ and $Z$ values, they found that the best fit model differed for each simulation. Furthemore, when comparing {\it identical} models (e.g., the exponentially tapered power law), the best fit parameters were different at a statistically significant level.

Taken together, these results call into question the originally hypothesized universality of the initial mass distribution. However, further work is required to determine what trends emerge with varying physical parameters.  Furthermore, a more rigorous analysis is called for in which the same tools (e.g., advanced clump-finding with {\sc Plan} and improved statistics via Markov-Chain Monte Carlo techniques) are applied consistently across a large region of parameter space.

Despite this complexity, there are some consistent features that emerge from these calculations. The slope of the mass distribution towards smaller masses generally falls within the range 1--2 (though not in all models; see \citealt{Li2019}) and thus the mass function is dominated in number by small planetesimals but in mass by large planetesimals (see Figure 4 of \citealt{Li2019}). Furthermore, the mass distributions are in the broadest sense composed of a low mass region (or regions) with shallow slopes and higher mass regions that drop more rapidly with mass, either in the form of an expoential cut-off or a steeper power law.

Of course, the ultimate question is what do these results tell us, if anything, when compared to observations? This is a point to which we will return shortly. However, as these results only address the {\it initial} planetesimal mass distribution, we must briefly discuss the role of post-formation collisional effects and how these may affect our understanding of the mass and size distributions of planetesimal populations.

\subsubsection{Collisional Evolution}
\label{sec:collisional}

There are two main stages to the collisional evolution of comets and KBOs.
During the first stage, bodies presumably resided in a massive ($\sim 20$ $M_\oplus$) trans-Neptunian disk (\jbs{see Chapter 4 by Kaib \& Volk, Chapter 5 by Fraser et al. and} \citealt{Morbidelli2020} for further discussion of KBO dynamics) and experienced strong
collisional grinding over the lifetime of the disk of solids ($t_{\rm s,disk}$).
They were subsequently implanted
into the Kuiper belt, scattered disk, and Oort cloud when Neptune migrated to 30 au (e.g., \citealt{Dones2015}). The collisional evolution during the subsequent 4-plus Gyr, the second stage of collisional evolution, is thought to be less significant; at these times, these populations were distant and/or did not contain much mass \citep{Singer2019,Greenstreet2019,Morbidelli2021}.

The collisional evolution during the massive disk stage was studied in \cite{Nesvorny2018}. They
showed that the Patroclus-Menoetius (P-M) binary in the Jupiter Trojan population poses an important
constraint on the disk lifetime. This is because the longer the P-M binary stays in the disk, the
greater is the likelihood that its components will be stripped from each other (by impacts). This
indicates $t_{\rm s,disk}<100$ Myr \citep{Clement2018,deSousa2020,Morgan2021}.

Different studies assumed different initial size distributions and impact scaling laws to model the
collisional evolution of KBOs/comets \citep{Pan2005,Fraser2009,CampoBagatin2012}. For example, \cite{Nesvorny2019a} adopted the initial size distribution inspired
by the SI simulations (e.g., \citealt{Simon2017}), and impact scaling laws for weak ice, to demonstrate that the collisional grinding would remove very large bodies and produce the KBO size-distribution
break at $D \sim 100$ km \citep{Bernstein2004,Fraser2009}. If that is the case, most comets with $D<10$ km must be fragments of large planetesimals since in collisional grinding, the collisional lifetime increases with the size of the body \citep{Morbidelli2015}. The small survivors would have suffered multiple reshaping collisions \citep{Jutzi2017a}. Bi-lobed comets should either be the product of sub-catastrophic collisions \citep{Jutzi2017b} or fragments of catastrophic collisions \citep{Schwartz2018}.

Jupiter Trojans have a well defined cumulative power index $-2.1$ from below 10 to 100 km \citep{Emery2015}, which is very close to the equilibrium slope expected for
the gravity regime ($D>0.5$ km; \citealt{OBrien2003}). This represents a very important
constraint on the collisional evolution. Specifically, if the Jupiter Trojan size distribution was
established during the massive disk stage, the disk must have been dynamically cold (large impact
velocities would produced waves in the size distribution that are not observed; e.g., \citealt{Fraser2009,Kenyon2020}). It is also possible, however, that $t_{\rm s,disk} = 0$, which corresponds to immediately after gas dispersal,  in which case the
observed size distribution of Jupiter Trojans and KBOs above $\sim$10 km would be primordial
(and an important constraint on the formation processes). Even in this case, however, it is
difficult to avoid the collisional disruption of comet-size bodies ($D<10$ km; e.g., \citealt{Jutzi2017a,Nesvorny2019a}), unless comets are jocularly stronger that we think.

The first stage of evolution in the massive planetesimal disk, when most collisional grinding happened, only applies to bodies that formed in the massive disk below 30 au. This presumably includes most bodies in the dynamically hot populations of the Kuiper belt (hot classicals, resonant objects, etc.), and comets, making a direct comparison with SI planetesimal formation models difficult (as the SI simulations do not include collisional evolution). However, cold-classical KBOs are thought to
have formed in a low-mass planetesimal disk at $>40$ au \citep{Parker2010,Batygin2011,Dawson2012},
not significantly collisionally evolved during stage two (stage one does not apply to them), and thus likely
represent the most pristine planetesimals in the Solar System.  As such, we restrict our comparison to these objects in what follows.

\subsubsection{Comparison Between Models and Cold Classical KBOs}
\label{sec:size_dist_compare}

To best compare numerical models with observations of the cold-classicals \jbs{(a discussion of the other KBO populations is covered in Chapter 5 by Fraser et al.)}, we can rewrite the relevant parameters of our mass functions as equivalent size distribution parameters. Namely, the cumulative size distribution for bodies of diameter $D$ in the three-parameter exponentially tapered power law (which we choose based on its common use in the literature thus far) is

\begin{equation}
    \label{eqn:NRcum1}
    N(>D) \propto D^{-q+1} \exp \left[\left(-D/D_c\right)^{\gamma}\right]
\end{equation}

\noindent
where $q = 3p-2$, $D = 2\left(M_c/4\pi \rhos\right)^{1/3}$, and $\gamma = 3\beta$.

From the discussion in Section~\ref{sec:primordial_mass_dist} and using the above relation between $q$ and $p$, $q$ ranges from 1.9 in the tapered models of \cite{Abod2019} to 2.8 for the simple power law \citep{Simon2016b,Simon2017} and the tapered model of \cite{Johansen2015}.

On the observational side, \cite{Fraser2014} analyzed the H band magnitude distribution of KBOs by combining the results of a number of independent KBO surveys. In fitting their data to two separate power laws (joined at a ``break point") of the form

\begin{equation}
    \label{eqn:NRcum2}
    N(<H) \propto 10^{\alpha H}
\end{equation}

\noindent
they were able to determine $\alpha \approx 1.5$ for the largest planetesimals, translating to $q = 5\alpha+1 = 8.5$, and $\alpha \approx 0.38$ for small planetesimals, equating to $q = 2.9$. The magnitude at the break corresponds to a KBO diameter of 140~km.

More recently, \cite{Kavelaars2021} re-examined this problem with new data collected from OSSOS as well as previous observations. They found that the cumulative magnitude distribution is best fit with an exponential cut-off corresponding to diameters ranging from 80~km to 130 km (assuming an albedo of 0.15) and a slope for small objects (the data for which were collected from other sources) very similar to the \cite{Fraser2014} value.

\begin{figure*}[t!]
    \centering
    \includegraphics[width=1.\columnwidth]{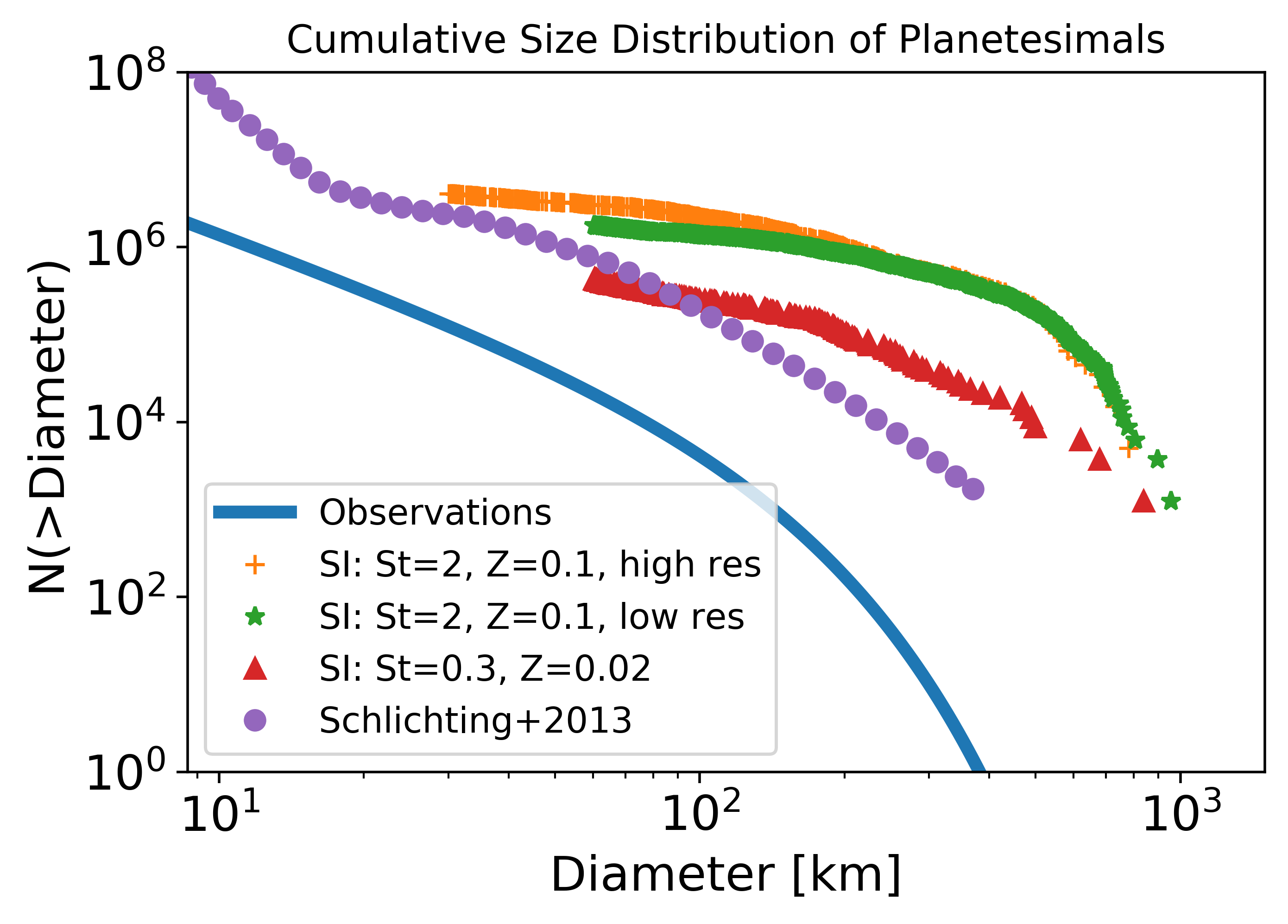}
    \includegraphics[width=1.\columnwidth]{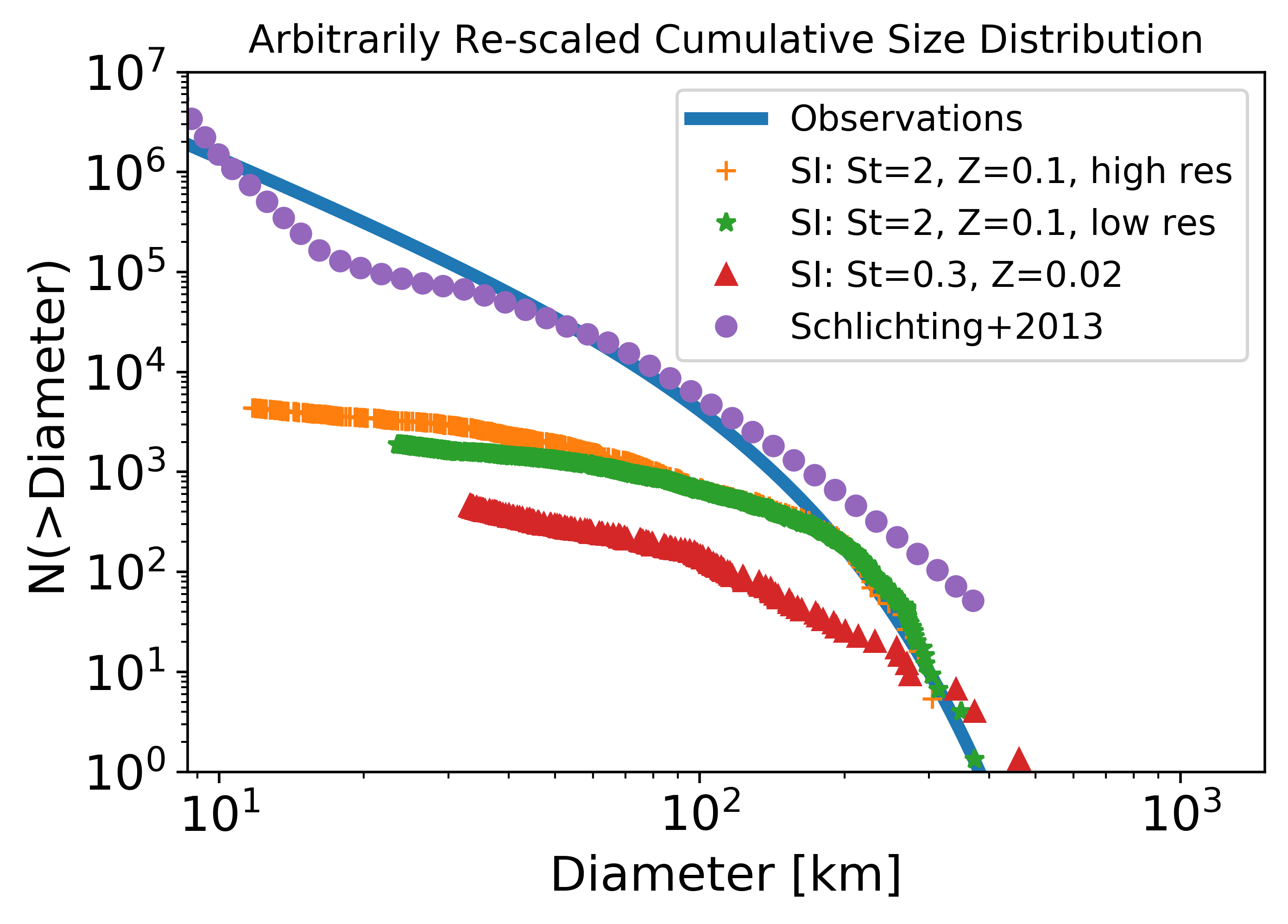}
    \caption{\jbs{A fit to the observed} cumulative size distribution of KBOs from the Outer Solar System Origins Survey (solid blue; \citealt{Kavelaars2021}) compared with theoretical predictions made by SI simulations with different parameters (orange pluses, green stars, red triangles) and from the gravitational coagulation model of \cite{Schlichting2013} (purple circles). \jbs{In the left plot, both the SI and coagulation calculations have been adjusted to account for the full extent of the Cold-Classical belt (e.g., we multiplied the data from \cite{Schlichting2013} by the projected area of the belt on the sky, which we assumed to be $180^{\circ}\times5^{\circ}$). The right plot shows the same data but with the models {\it arbitrarily re-scaled} so that at least some of the predicted distribution lines up with the data. For the SI simulations, $\St$, $Z$, and the resolution are varied. Neither the SI simulations, nor the coagulation model, match the observations, both in terms of numbers of planetesimals and the shapes of the distributions.}}
    \label{fig:size_dist_compare}
\end{figure*}

While the $q \approx 2.8$ from the numerical models is temptingly close to the value inferred from these observations, we must reiterate that $q = 2.8$ is only the best fit for simpler power laws (which are clearly not an accurate representation of planetesimal size distributions).  In fact, with the exception of the one simulation in \cite{Johansen2015},\footnote{While this simulation has a slope that agrees with observations at small sizes, there is significant mismatch between the simulations and observations at larger sizes (not shown in Fig.~\ref{fig:size_dist_compare}).} the models that are fit with an exponentially tapered power law (which at least agrees with the shape of the \cite{Kavelaars2021} fit in principle, though see below) have much smaller $q$ values.

Beyond the slope, one can also compare characteristic planetesimal sizes.
A fit to a characteristic diameter was not analyzed explicitly in \cite{Simon2016b} and \cite{Simon2017}, but the effective diameters of their largest planetesimals were $\sim 200$--800km. Similarly, \cite{Schafer2017} found large planetesimals with diameters $\approx$ 600--1200 km (when scaling their units to the Kuiper Belt), and the maximum planetesimal diameters of \cite{Li2019} ranged from 320~km to 1200~km.  On the other hand, \cite{Abod2019} found a narrower range of characteristic diameters: $\sim 200$--400~km, which is more consistent with observations, though still slightly larger than those inferred from observations.

\jbs{Finally, one should consider the total number of planetesimals, as the debiased data in \cite{Kavelaars2021} put a constraint on the total number of cold-classical KBOs.  When adjusting for the fact that local, shearing boxes represent only a very small fraction of the Cold-Classical belt, SI calculations generally produce too many planetesimals. For instance, the  $\St = 0.3, Z = 0.02$ model of \citep{Simon2017} produces a factor of $\sim$ 10 more planetesimals (at least at the resolution probed by the simulations) compared to the OSSOS data. }

To summarize the take away points of this comparison, Figure~\ref{fig:size_dist_compare} shows the cumulative size distribution from three SI simulations representing different regions of numerical and physical parameter space \jbs{as well as a fit to the observed data as described in} \cite{Kavelaars2021}. \jbs{Clearly, there remains a mismatch between the size distribution shape, the total number of planetesimals, and the maximum (or characteristic) planetesimal size. Though, as shown in the right plot of Fig.~\ref{fig:size_dist_compare}, the slope of the SI simulations with $\St = 2, Z = 0.1$ do match the observations reasonably well when the theoretical curves have been manually rescaled to lie on top of the observational data.}

These discrepancies between theory and observation are not yet resolved. One possibility \jbs{regarding planetesimal sizes and the shape of their size distribution is that due to pebble clouds} not collapsing below a numerically limited spatial scale in the {\sc Athena} simulations (see above), such clouds may accrete pebbles at an unrealistically large rate or suffer interactions with each other (e.g., tidal stripping, merging).  However, this issue is not present with the sink particle runs of {\sc Pencil}, and there are still discrepancies between those simulations and KBO observations (e.g., \citealt{Schaefer2017}).

Another possibility is that since the continued collapse of the pebble clouds leads to the formation of multiple smaller planetesimals, as discussed in Section~\ref{sec:final} and \cite{Nesvorny2021}, the initial mass/size distributions will change. While the fact that many of these clouds put most of their mass into a binary seems to suggest that the distributions will not change appreciably, more work is required to resolve this issue.  In particular, a larger exploration of parameter space (both for the SI and for the relevant parameters in the {\sc Pkdgrav} calculations) is required, as well as a direct calculation of the size distribution from the final collapsed planetesimals.

\jbs{
     It is also possible that a larger volume of the SI parameter space requires exploration.  Even with the many publications on the SI mass function, there remain many parameter combinations that have not been explored.  For the sake of computational expediency, most of the SI simulations have been carried out far from the border between strong clumping and no strong clumping (see Fig.~\ref{fig:strong_clumping}). In reality though, as pebbles grow from smaller grains, regions of the disk may move from left to right in the $Z$-$\St$ parameter space, and thus the relevant $\St$ values may be smaller than what has been studied so far. While, to the best of our knowledge, small \St values have only been explored within the context of the size distribution in \cite{Simon2017}, a less sophisticated clump-detection method in that work led to poorer sampling of planetesimals, ultimately resulting in a simple power law fit to the data (as discussed above). Similarly, the $Z$ values may be closer to the border and not as high as $Z = 0.1$, unless there is a sudden rapid increase in the dust-to-gas ratio in the disk such that the system is rapidly placed well within the strong clumping regime. 
    
    Indeed, simulations along this border have shown some indication of marginal behavior (e.g., transient excursions into the strong clumping regime; \citealt{Li2021}). If true, then this would be a natural route towards reducing the number of planetesimals formed by the SI by e.g., limiting the amount of time or spatial regions in which strong clumping occurs. How exactly this marginal behavior would affect the size distribution and maximum planetesimal sizes is less clear, but certainly remains a question worthy of study.
    
    Finally, it is also possible that including a range of particle sizes will change the results, given the role that multiple particle sizes play in the linear regime of the SI \citep{Krapp2019,Paardekooper2020,Zhu2021}. However, as discussed above, simulations of the non-linear state of the SI that include the vertical component of stellar gravity suggest that the nature of strong clumping does not change drastically change when multiple particle sizes, at least not to the extent suggested by linear analyses \citep{Bai2010_3D,Schaffer2018,Schaffer2021}. 
    
}

The coagulation models for planetesimal formation also make specific predictions for the size distribution of these bodies.  \cite{Schlichting2011} predict that $q \approx 4$ over a large range of planetesimal sizes, which is largely inconsistent with the most recent observational constraints. An updated model that includes collisional evolution \citep{Schlichting2013} (shown in Fig.~\ref{fig:size_dist_compare}) also does not agree with observations.  \jbs{As with the SI, there are too many planetesimals compared with that from \cite{Kavelaars2021}.}
In particular, this model produced an increasingly large number of $D < 2$~km bodies (not shown on the graph).  This excess of small bodies is inconsistent with recent measurements of cratering records on Pluto and Charon showing a turn over in the size distribution \citep{Singer2019} as well as even more recent work that, while suggesting a turn over does not exist, limits the slope to being shallower than predicted \citep{Morbidelli2021}. Beyond this, the general shape of the \cite{Schlichting2013} size distribution is inconsistent with observations, as shown in the figure.

The numerical calculations of \cite{Kenyon2012} more broadly demonstrated that a wide range of size distribution slopes $q$ (their results had different $q$ values for different planetesimal size ranges) were possible depending on their initial conditions and chosen parameters. However, they generally found that $q \approx 3$ for $D > 200$km; again, this is inconsistent with observations. However, it is worth noting that at smaller planetesimal diameters, $q \approx 3$--3.5, which {\it is } approximately consistent.

\jbs{Finally, the turbulent clustering mechanism  makes specific predictions for the initial size distribution of planetesimals. In particular, as outlined in \cite{Hartlep2020}, the initial size distribution has a well defined peak of between 10 and 100 km in diameter, depending on the relevant model parameters\footnote{\jbs{The total number of planetesimals formed via this mechanism is also largely dependent on these parameters \citep{Hartlep2020}.}}, and does not have a power law shape at smaller scales.  While the characteristic size of planetesimals formed via this mechanism is of approximately the same order of magnitude as observed objects, there is still disagreement in the shape of the distribution compared with observations.}

In summary, the various gravitational collapse mechanisms that have quantified the size distribution as well as the coagulation model have yet to completely reproduce the typical sizes of planetesimals, their size distributions, and the total number of bodies formed, strongly motivating future work in this area. However, as we will now see, another diagnostic {\it has} proven to be a powerful discriminant between two of these models.

\subsection{Constraints from KBO Binaries}

\begin{figure}[t!]
    \centering
    \includegraphics[width=1.\columnwidth]{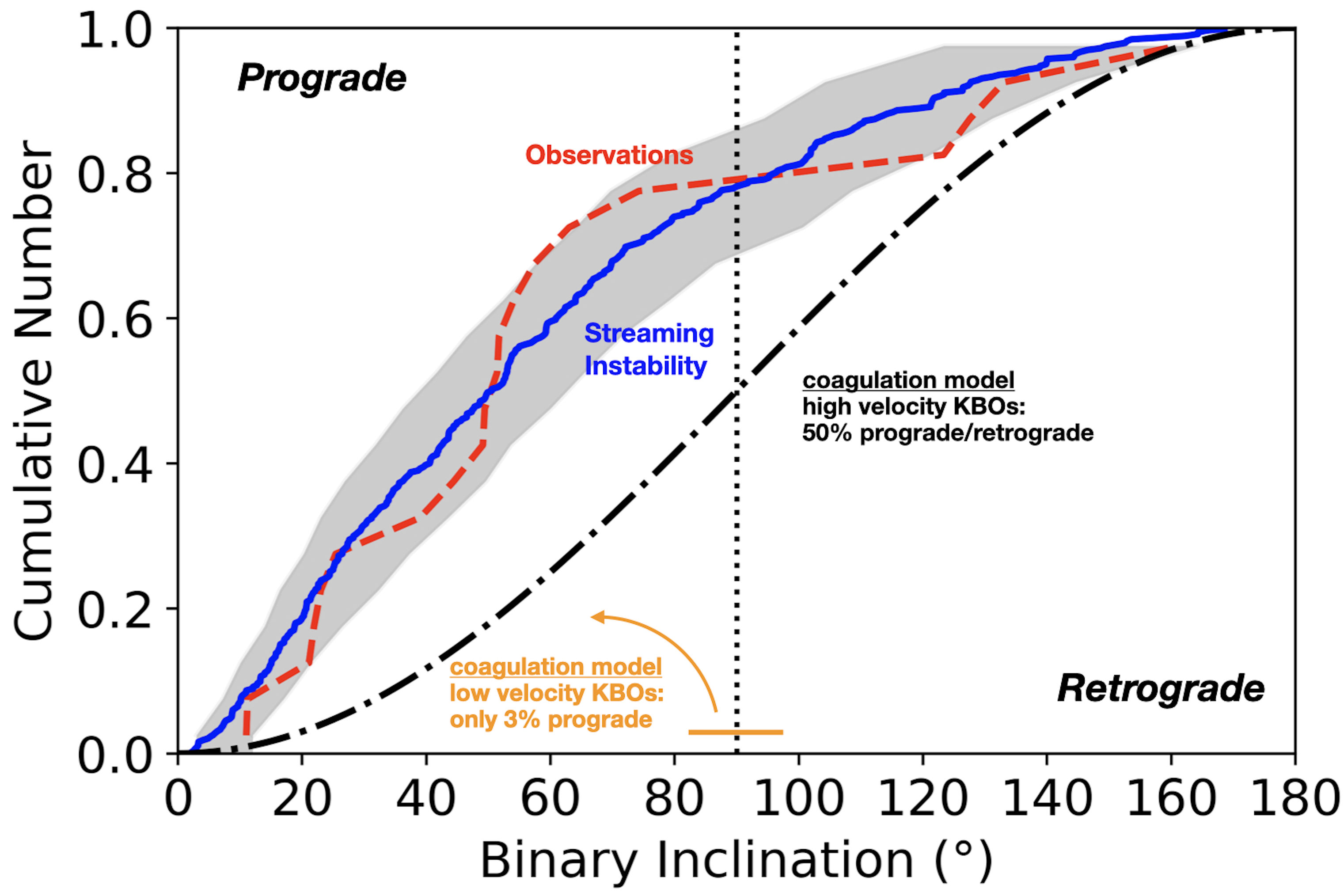}
    \caption{The inclination distribution of binary orbits in a snapshot of the $\St = 2, Z =0.1$, higher resolution run of \citep{Nesvorny2019} (blue solid line) and observations of binary KBOs (red dashed line).  The shaded area is a 68\% envelope
        of the model, which results when a sample of model orbits is randomly drawn from a sample size equal to that of real binaries with known inclinations. The binary orbits are predominantly prograde with approximately 80\% having inclination < 90°. The competing gravitational coagulation model predicts either a 50\% split between prograde and retrograde (black dash-dot line, which was calculated assuming a uniform distribution of inclinations) or 3\% prograde and 97\% retrograde (see orange horizontal marker).  The SI model is in excellent agreement with observations, in contrast to the coagulation model.
    }
    \label{fig:binary_compare}
\end{figure}

It is now known that many observed KBOs are actually in a binary configuration (e.g.,  \citealt{Noll2008}). More recently, \cite{Grundy2019} analyzed 35 such binaries and found that $\approx 80\%$ of them orbit each other in a prograde sense.

\cite{Nesvorny2019} showed that the SI mechanism produces {\it a nearly identical} distribution of binary orbital orientations: 80\% of the simulated planetesimals orbit prograde.  This astonishingly strong agreement is shown in the cumulative distribution (see Fig.~\ref{fig:binary_compare}) of orbital inclination for both SI models and observations.  Even more encouragingly, \cite{Nesvorny2021}\footnote{Also, see \cite{Nesvorny2010} and \cite{Robinson2020} for similar work on pebble cloud collapse} extended this and found that the angular momentum {\it magnitude} of SI-produced binaries also agrees with observed KBO binaries.

Future work to test the SI hypothesis for planetesimal formation includes a further exploration of parameter space as well as extending to a larger number of simulated particles in {\sc Pkdgrav}. The latter is important to numerically resolving the tight and contact binaries, such as Arrokoth \citep{McKinnon2020}.

In contrast, the gravitational coagulation model described in Section~\ref{sec:size_dist_compare} makes predictions that are inconsistent with these observations.  More specifically, if the KBO velocities during this coagulation stage are on the order of or greater than the so-called ``Hill velocity" (see \citealt{Schlichting2008} for a definition), this model predicts roughly equal number of prograde and retrograde orbits (this line is also included in Fig.~\ref{fig:binary_compare}). For KBO velocities less than 10\% of the Hill velocity, retrograde orbits dominate by 97\% \citep{Schlichting2008}.   Clearly, these considerations taken together strongly support the SI-induced gravitational collapse paradigm.

\subsection{\label{sec:67p} Comparison with Comet 67P}

Moving from constraints from KBOs to those from comets, we now consider recent observations of comet 67P Churyumov-Gerasmineko (hereafter comet 67P) and compare these with model predictions of planetesimal-formation and -evolution models. This comparison will include internal properties, like porosity, granularity, cohesive and tensile strength, and activity patterns.

Owing to the ground-breaking {\it Rosetta} mission of the European Space Agency, our knowledge about the inner makeup of cometary nuclei has increased enormously. We are now in the situation to use these findings as diagnostic tools for the formation of these bodies \citep[see][]{Blum2017,Blum2018,Weissman2020}. The most striking physical property of comets for which we now possess empirical evidence is their extremely low mechanical strength, claimed to be essential for the sustaining activity of comets \citep{Kuehrt1994,Skorov2012,Blum2014}. Measurements of the tensile strength of the cometary material have shown that the tensile strength indeed is $\lesssim 1$~Pa \citep{Attree2018}. Numerical studies suggest that such a requirement is mandatory for a sustained cometary activity \citep{Skorov2012,Gundlach2020}. Extremely low compressive-strength values were also inferred from the bouncing of the Philae lander \citep{ORourke2020}.

{\it Rosetta} also allowed precise mass and volume determinations of the nucleus of comet 67P \citep{Paetzold2019}. In combination with assumptions on the material composition, this allows the derivation of global porosity values. Different approaches result in values for the porosity of 73\%–76\% \citep{Herique2019}, 65\%–80\% \citep{Paetzold2019}, and 63\%–79\% \citep{Fulle2016}, which correspond to global volume filling factors of the nucleus of comet 67P of $\Phi_\mathrm{67P}=0.24-0.27$, $\Phi_\mathrm{67P}=0.20-0.35$, and $\Phi_\mathrm{67P}=0.21-0.37$, respectively. As pointed out in Sect. \ref{sec:final}, the expected volume filling factor for small ($\lesssim 100$~km) planetesimals is $\Phi_\mathrm{pl} \approx 0.18-0.37$, which is well matched by the values estimated for comet 67P. \citet{Guettler2019} showed that such volume filling factors indicate a makeup consisting of two hierarchical levels (e.g. solid dust grains and fluffy pebbles). Fewer levels lead to too high densities, whereas more levels render the planetesimal too porous.

A "smoking gun" for the primordial nature of comet 67P could be the discovery of extremely fluffy dust particles by the Giada and Midas instruments of {\it Rosetta} \citep[see the review by][and original references therein]{Guettler2019}. These particles possess porosities $>95\%$ and \jb{have pre-detection sizes on the order of 1 cm, but electrostatically fragment in the vicinity of the Rosetta spacecraft. Even under the most optimistic conditions, such particles could not have formed in the coma of comet 67P. To see this, consider the  maximum measured dust mass flux of comet 67P: $F=10^{-8} \, \mathrm{kg\, m^{-2}\, s^{-1}}$ \citep{Merouane2016}. This result, along with the assumption that the monomer grains have sizes of $\sim 100$\,nm (see Sect. \ref{sec:st-bo-fr}) and masses of $m=10^{-18}\, \mathrm{kg}$, a particle flux (per unit mass) of $F^{\prime} = F/m = 10^{10}\, \mathrm{ m^{-2}\, s^{-1}}$ results. The dust monomers are so small that they couple to the gas outflow on short timescales and, thus, obtain typical outflow velocities of $v_\mathrm{out}=100 \, \mathrm{m\, s^{-1}}$. Hence, a maximum number density of the monomer grains in the coma of $n= F^\prime/ v_\mathrm{out}= 10^8 \, \mathrm{m^{-3}}$ results. To determine the collision time scale, we assume that all collisions occur at the sticking-bouncing threshold, ignoring that fractal growth requires much lower collision speeds. From the models described in Sect. \ref{sec:st-bo-fr}, the sticking-bouncing threshold velocity of 100 nm grains should be on the order of $v_\mathrm{r}=10 \, \mathrm{m\, s^{-1}}$. With a collision cross section for 100 nm grains of $\sigma \approx 10^{-13} \, \mathrm{m^2}$, a collision timescale of $\tau_\mathrm{coll} = \frac{1}{n \sigma v_\mathrm{r}} = 10^4 \, \mathrm{s}$ follows. At a distance of Rosetta from the surface of comet 67P of $d= 10-30 \, \mathrm{km}$ \citep{DellaCorte2015}, the time between emission of the dust and detection by the Rosetta/Giada instrument was $\tau_\mathrm{trans} = d / v_\mathrm{out}= 100-300 \, \mathrm{s}$, much too short to allow any collisional growth of the grains. Thus, the observed fluffy grains must have been present inside the nucleus of comet 67P before they were ejected into the coma.}

\citet{Fulle2017} showed that these fluffy particles could be the remnants of the pre-pebble growth stage of protoplanetary dust, captured and preserved between the pebbles during the gravitational collapse. If that scenario was correct, the cometary nuclei in the Solar System could not be rubble piles, re-assembled after a catastrophic collision. If post-formation collisions were unavoidable, then they had to be sub-catastrophic so that major parts of the colliding planetesimals would remain intact \citep[see][for an example]{Schwartz2018}.

Under the assumption that comet 67P was formed by the gentle gravitational collapse of a pebble cloud, \citet{Blum2017} used a multi-Rosetta-instrument approach to estimate the pebble sizes. They found that pebble radii in the range of 3-6~mm were the most likely, with the upper limit constrained by the sub-surface temperatures measured by the Miro instrument and the lower limit given by the size distribution of surface features identified by the Civa instrument on the Philae lander. These pebble sizes are in good agreement with estimates of the bouncing barrier at large heliocentric distances \citep[see][and Sect. \ref{sec:sizedis}]{Lorek2018}.

We should note that there are alternative formation models for the nucleus of comet 67P \citep[see Sect. \ref{sec:coagmod} and][for a recent discussion on this issue]{Weissman2020} and that there are open questions with the formation of cometary nuclei by the gentle gravitational collapse of a pebble cloud. The most prominent issues to be solved in future work are: (i) What is the influence of radiogenic heating on the inner makeup of planetesimals? (see Sect. \ref{sec:final}) (ii) Can the gravitational collapse form comet-sized planetesimals directly and what is their rate of binarity? (iii) What is the role of collisions in the post-formation era and how can present-day cometary nuclei form as a result of these collisions without losing their volatile and fluffy-particle \jb{inventory}?
That being said, the evidence for a gravitational collapse origin of cometesimals is reasonably compelling, giving further support to the SI as the mechanism to form planetesimals (though of course, other gravitational collapse mechanisms cannot be ruled out).

\section{CONCLUDING REMARKS}

Less than two decades ago, the formation of planetesimals was one of the biggest unanswered questions in planetary science and astrophysics. While this issue does remain open and of utmost importance to understanding the formation of the Solar System (as well as other planetary systems), it is very encouraging that so much progress has been made in the time since then.

We now have a detailed understanding of how sub-micron grains grow to pebble sizes, backed by both theoretical calculations and laboratory experiments, and we have {\it multiple} models to explain how pebbles then reach the next stage and form planetesimals.

Of course, there remain issues that must be solved in order to build towards a comprehensive understanding of and to determine which mechanisms are responsible for planetesimal formation. For instance, the largest issue of contention is the discrepancy in the size distribution between observations and theoretical models, those based on both the SI and gravitational coagulation. Will other gravitational collapse mechanisms (e.g., the SGI) show better agreement? Or is the discrepancy, at least in the SI simulations, the result of something simpler, such as \jbs{a more extensive survey of relevant parameters} or collapse below the numerically limited scales in the SI calculations?

Even with these outstanding issues, however, there are indications that the gravitational collapse paradigm, and specifically the SI-induced collapse model, is the correct one.  The thermal and structural properties of planetesimal made available from the {\it Rosetta} mission do indeed support a gravitational collapse model for planetesimal formation.  Furthermore, the excellent agreement in KBO binary properties between SI simulations and observations {\it strongly} support the SI as the mechanism for planetesimal formation, while strongly disfavoring the gravitational coagulation model.

Furthermore, as numerical models improve and computational resources continue to be readily accessible, the theoretical side of this question will likely see significant progress in the coming years.  These enhancements will be further augmented by improved observational capabilities. For example, in addition to continued observations from OSSOS, the Vera C. Rubin observatory will drastically expand our database of detected KBO objects, while {\it JWST} will provide another way to peer into the earliest stages of planet formation in our Solar System. \jbs{Furthermore, new missions, such as {\it Lucy} and {\it Psyche}, are either being built or are already on their way to exploring Solar System planetesimals.}

In closing, with ever improving numerical models, new observational facilities coming online, and new missions set to explore Solar System planetesimals, we live in a golden era of planetesimal and comet exploration. As such, there should be significant optimism that we as a community will solve the outstanding issues of planetesimal and comet formation in the coming years; the future of small body studies looks quite bright, indeed.

\vskip .5in
\noindent \textbf{Acknowledgments.} \\
J.B.S. acknowledges support from NASA under {\em Emerging Worlds} through grant 80NSSC21K0037. J.B. acknowledges the support of this work by the Deutsche Forschungsgemeinschaft (DFG) through grant BL 298/27-1 and thanks for continuous support by DFG and the Deutsches Zentrum f\"ur Luft- und Raumfahrt \jb{-- German Space Agency}. T.B. acknowledges funding from the European Research Council (ERC) under the European Union’s Horizon 2020 research and innovation programme under grant agreement No 714769 and funding by the Deutsche Forschungsgemeinschaft (DFG, German Research Foundation) under Ref no. FOR 2634/1 and under Germany's Excellence Strategy - EXC-2094 - 390783311. D.N. acknowledges support from the NASA Emerging Worlds program. We thank Daniel Carrera, Wladimir Lyra, Andrew Youdin, and Rixin Li for useful discussions that greatly improved the quality of this chapter.  Furthermore, we thank Jono Squire for providing the necessary materials and information to create Fig.~\ref{fig:SI_mechanism}. We also thank Ingo von Borstel for creating the fractal aggregate in Fig. \ref{fig:fractals}. \jbs{Finally, the writing of this document was supported in part by the Munich Institute for Astro- and Particle Physics (MIAPP) which is funded by the Deutsche Forschungsgemeinschaft (DFG, German Research Foundation) under Germany´s Excellence Strategy – EXC-2094 – 390783311, and} J.B.S. thanks MIAPP for their hospitality while he finished much of this chapter.

\end{document}